\documentclass[twocolumn,nofootinbib,preprintnumbers,amsmath,amssymb]{revtex4}
\usepackage{graphicx}
\usepackage{epsfig}
\newcommand{\be}{\begin{equation}}
\newcommand{\ee}{\end{equation}}
\newcommand{\ba}{\begin{eqnarray}}
\newcommand{\ea}{\end{eqnarray}}

\begin{document}
\title{Compactified String Theories - Generic Predictions for Particle Physics}\thanks{Invited Review for {\it International Journal of Modern Physics A}}
\author{Bobby Samir Acharya$^{1,2}$}
\author{Gordon Kane$^{3}$}
\author{Piyush Kumar$^{4}$}
\affiliation{$^1$ International Centre for Theoretical Physics, Trieste,  Italy \\ \\$^2$Department of Physics, King's College London, UK
\\ \\$^3$Michigan Center for Theoretical Physics, Ann Arbor, MI 48109 USA \\ \\
$^4$Department of Physics, Columbia University, New York, NY 10027 USA}

\begin{abstract}
In recent years it has been realized that in string/$M$ theories compactified to four
dimensions which satisfy cosmological constraints,  it is possible to make some generic predictions for particle physics and dark matter: a non-thermal cosmological history before primordial nucleosynthesis, 
a scale of supersymmetry breaking which is ``high" as in gravity mediation, scalar superpartners too heavy to be produced at the LHC (although gluino production is expected in
many cases), and a significant fraction of dark matter in the form of axions. When the matter and gauge spectrum below the compactification scale is that of the MSSM, a robust prediction of about 125 GeV for the Higgs boson mass, predictions for various aspects of dark matter physics, as well as predictions for future precision measurements, can be made. As a prototypical example, $M$ theory
compactified on a manifold of $G_2$ holonomy leads to a good candidate for our ``string vacuum", with the TeV scale emerging from the Planck scale, a de Sitter
vacuum, robust electroweak symmetry breaking, and solutions of the weak and
strong CP problems. In this article we review how these and other results were derived, from the key theoretical ideas to the final phenomenological predictions.
\end{abstract}
\maketitle
\newpage
\vspace{-1.2cm}
\tableofcontents

\section{Introduction}\label{intro}

Particle physics is currently entering an exciting era for a number of reasons. Significant data is being published by the CERN Large Hadron Collider (LHC) experiments as well as from dark matter searches. Well appreciated is the fact that this data will have profound implications for our understanding of electroweak symmetry breaking (EWSB) 
and the gauge hierarchy problem. Furthermore, this data could also have significant implications 
for physics at very high scales, such as the Grand Unified Theory (GUT) scale  or the string scale. 
 
There is a second reason why we are entering an exciting era. Developments over the past few years have led us to an understanding that, under some very general, simple and broad assumptions, 
string/$M$ theory provides a
\emph{framework} that (in practice) is capable of addressing key, fundamental questions about particle physics and cosmology. Moreover, the framework
addresses them in a unified way. 
At first sight, this may appear to be a surprising statement since we presumably still have a lot to learn about string theory.
Furthermore, there are an enormous number of solutions to string/$M$ theory which describe, effectively four-dimensional Universes  -- 
the string landscape. 
The enormity of the landscape has led to the popular, but incorrect, 
view that string theory has no predictive power and virtually \emph{any} low-energy theory could be a part of the landscape.

In this article we will review some of the results
which demonstrate that, on the contrary, 
\emph{generic predictions} do arise from string/$M$ theory which can be directly tested at current and future particle physics experiments as well 
as with astrophysical and cosmological observations. One of the main purposes of this review is to present these predictions from string/$M$ theory in
a unified fashion in a single, relatively short, fairly non-technical document.

The following section contains a summary of the basic ideas and results and can be read independently from the rest of the article.
The remainder of the article reviews the results in more detail.

\section{Summary of Basic Idea and Generic Predictions}\label{summary}

In recent years, investigations into string/$M$ theory based on an improved understanding of moduli and axion dynamics
have lead to a consistent, simple picture for physics beyond the Standard Model addressing collider phenomenology and dark matter.
This article is aimed at collecting and reviewing these results in a single document. This section is a very short, fairly self-contained
summary of the basic ideas and results and can be read independently from the remainder of the article.

Throughout this article, the basic assumption we make is that our 
Universe is described by a solution of string/$M$ theory.  In order to 
test this hypothesis we elicit from the theory the simplest, generic 
consequences which could describe our Universe and are relevant for particle physics experiments. We  
focus only on string/$M$ theory solutions with low energy supersymmetry and grand unification (at around\footnote{We will use `natural' units in which
$c = \hbar = 1$} $10^{16}$ GeV).

Given the above, the physics below the GUT scale can be effectively described by a four dimensional supergravity theory whose field content is at least that of the minimal supersymmetric Standard Model (MSSM) \cite{Dimopoulos:1982af}. This four dimensional theory can be thought of as any other quantum field theory, but with an essential difference. 
The theory also contains moduli and axion fields, which parametrize the size and shape of extra dimensions, as well as  their couplings to matter and to each other. The moduli and axions are essentially the only low energy remnants of the string/$M$ theory origin of the effective theory, and all other string/$M$ theory modes are decoupled from the four dimensional theory. There are typically large numbers of moduli and axion fields and, moreover, the axion decay constants are of order the GUT scale \cite{GSW}. We would now like to ask: if we consider a {\it generic}
solution of string/$M$ theory with low energy supersymmetry and grand unification, what phenomena does it describe? This is completely analogous to asking within the framework of
quantum field theory, for example:
what are the generic predictions of chiral gauge theories with hierarchical Yukawa couplings and spontaneous symmetry breaking? Essentially, if we threw a dart at the set of 
all solutions of string/$M$ theory which reduce to the Standard Model for physics processes below the TeV scale, what would the properties of that solution be?

The key to answering this question lies in the physics of the moduli and axion fields and the effective supergravity theory. In the supergravity theory, the mass of the gravitino (the
superpartner of the graviton),
$m_{3/2}$, sets the scale for the masses of {\it all scalar fields} unless symmetries prevent this. This is borne out by explicit string/$M$ theory calculations.
One of the key results that underlies many of the  predictions is a 
connection between the lightest moduli mass and the gravitino mass.
Essentially, the gravitino mass $m_{3/2}$ is related to the lightest modulus
mass (the smallest eigenvalue of the extended moduli mass matrix) by an ${\cal O}(1)$ factor.  
Details of the derivation are given in \cite{Acharya:2010af} and in section \ref{mod-spectra}.

In fact, both the MSSM scalars and the moduli fields will have masses of order $m_{3/2}$.
This is not true of the axion fields $a_i$ due to the shift symmetries $a_i \rightarrow a_i + c_i$ that originate from gauge invariance in higher dimensions. The moduli fields have couplings to matter which are suppressed by the Planck
scale. Therefore, they do not thermalize in the early Universe after inflation. Instead, when the Hubble scale decreases to become of order their mass $\sim m_{3/2}$, they quickly
come to dominate the energy density of the Universe, coherently oscillating in their potential. 
The lifetime of the moduli, which is generically of order ${m_{pl}^2 \over m_{3/2}^3}$ must be shorter than the age of the Universe at the onset of 
big-bang nucleosynthesis (BBN); otherwise, the moduli decay products, which include hadrons and leptons, would have a dramatic impact on the successful predictions of BBN. 
This is the infamous cosmological moduli problem \cite{CMP}.
This requires $m_{3/2} \geq$ 30 TeV or so\footnote{The answer is tens of TeV. 30 TeV is just a benchmark value.}. 
As we will see momentarily, axion physics provides a strong motivation
for $m_{3/2}$ to be as close to the BBN limit as possible. We thus take $m_{3/2} \sim $ 30 TeV implying the moduli dominate the energy density of the Universe right up to
BBN. The pre-BBN Universe is thus {\it matter dominated} and {\it not radiation dominated} as is often assumed in particle physics models. This is another generic prediction of
string/$M$ theory with low scale supersymmetry.

Now consider the axions. Due to the shift symmetries mentioned above, there are no perturbative contributions to their potential. Non-perturbative effects though, such as 
strong gauge dynamics, gauge instantons, gaugino
condensation and stringy instantons will generate a potential for the axions. Because any such contribution is exponentially suppressed by couplings and/or extra-dimensional volumes, in
our world with perturbative gauge couplings (at high scales), the axion masses will be exponentially small. Furthermore, since there are large numbers of axions in general, their masses will essentially be uniformly 
distributed on a logarithmic scale \cite{Arvanitaki:2009fg}. See \cite{Acharya:2010zx} for a detailed calculation of axion masses in string/$M$ effective theories. 
Like the moduli, the axions are also very weakly coupled to matter
and therefore do not thermalize in general. Moreover, since their masses are tiny - ranging from $m_{3/2}$ to even below the Hubble scale today - many of them, including the
QCD axion, start coherent oscillations during the time that the moduli are dominating the energy density, but before BBN. 

When the moduli decay, they release a large amount of entropy, and this dilutes the energy density of all species by a large amount
(ten orders of magnitude is a typical number to have in mind). 
This dilution has several important consequences. In particular, the energy density of any dark matter candidates present during this epoch will be diluted
dramatically. This applies to both WIMPs and the axions themselves, implying that the bounds on the WIMP cross-section and axion decay constants coming the observed
dark matter density assuming a radiation dominated early Universe are not applicable. In the case of axions the bound on the axion decay
constant increases dramatically from 10$^{12}$ GeV in the radiation dominated Universe 
to 10$^{15}$ GeV in the string/$M$ theory case without fine-tuning \cite{Fox:2004kb,Kaplan:2006vm,Acharya:2010zx}. Therefore, with a small amount of tuning (1-10\,$\%$), axions with GUT scale decay constants are cosmologically consistent and will constitute a significant fraction of dark matter today. This is another generic prediction.
Note that, increasing the
gravitino mass increases the fine-tuning in the axion relic density \cite{Acharya:2010zx}, so 30 TeV is roughly the preferred value for $m_{3/2}$. 
Therefore {\it the solution to the cosmological moduli problem and axion physics both set the scale of supersymmetry breaking (as
characterized by $m_{3/2}$) to be of order 30 TeV.}
We will return to discuss WIMPs in this context after discussing the masses of
supersymmetric particles.

In a generic supergravity theory, the masses of all canonically normalized scalars (except axions) will be close to $m_{3/2}$. So, with $m_{3/2} \sim $ 30 TeV, the supersymmetry breaking mass parameters (at the GUT scale) of all scalars are also of order 30 TeV. This is confirmed by explicit string/$M$ theory calculations. In realistic cases with moduli stabilization it can also be shown that the masses of the SM-charged scalars, schematically denoted by $m_{soft}$, are: \ba\label{scalar} m_{soft}^2=m_{3/2}^2\,(1+{\rm small\,corrections}) \ea (see section \ref{spectra} for more details.) Even after renormalization to the weak scale, this implies that scalar squarks and sleptons will not be produced directly at the Large Hadron Collider (LHC). Quasi-degenerate squark and slepton masses at this scale are consistent with flavor observables, such as flavor-changing-neutral-currents (FCNCs).
However, a full solution of flavor issues within this framework requires more understanding. With $\sim 30$ TeV scalars, however, the supersymmetric (weak) CP problem, on the other hand, can be naturally solved for the class of moduli and axion stabilization mechanisms which also solve the strong CP problem. This is essentially because for such cases, the superpotential and soft terms in the vacuum are effectively {\it real} at leading order. Moreover, an improvement of a few orders of magnitude in the sensitivities of future EDM experiments could further test the framework.

While the squarks and sleptons have masses of order $m_{3/2}$,
the same is not necessarily true of the fermionic superpartners -- the gauginos and higgsinos. This could be either due to {\it symmetries} \cite{Wells:2004di} or {\it dynamics}. As an example of the latter, this happens if the field which dominates
supersymmetry breaking is not a modulus. In this case it does not contribute to gaugino masses since the gauge couplings are linear combinations of moduli fields in string/$M$ theory.. Many string theory examples with such properties exist in the literature. For example,  in the $G_2$-MSSM model which arises from $M$ theory \cite{arXiv:0801.0478}, supersymmetry breaking is dominated
by a hidden sector matter field. The suppression of the masses turns out to be one-to-two orders of magnitude and is parametrically of
order ${\alpha_h \over 4\pi}$, where $\alpha_h = {g_h^2 \over 4\pi}$ and $g_h$ is a hidden sector gauge coupling. 
Furthermore, if the doublet-triplet splitting problem is solved by a discrete symmetry in $M$ theory as proposed by Witten \cite{Witten:2001bf}, then 
there is also a suppression of the higgsino masses \cite{arXiv:1102.0556}, though this suppression is weaker than that of the gauginos in detail. We will consider both suppressed and un-suppressed gaugino masses and higgsino masses in what follows, though we will see that the suppressed case is favored both theoretically and phenomenologically.

If gaugino masses are suppressed, then they will be produced in the early Universe when the moduli decay. In particular, if the lightest supersymmetric particle (LSP)
is stable, it will also constitute a component of dark matter, in addition to the axions. One can estimate the LSP relic density \cite{Moroi:1999zb, Acharya:2008bk} and one finds that an 
LSP with an annihilation cross-section comparable to that of a neutral wino with a mass of order 200 GeV has an abundance of order the observed relic density today, 
resulting in a `non-thermal WIMP miracle'
\cite{Acharya:2009zt}. Furthermore, in this case, gauginos can be directly produced at the LHC. The main signal there is the direct production of gluinos. Such gluinos must
decay through intermediate off-shell squarks. Since third-generation renormalization effects tend to make the stop squarks the lightest of the squarks, the gluino has a sizable
branching fraction into top or bottom quarks via $\tilde{g} \rightarrow t\bar{t} \chi_1^{0}$ or $\tilde{g} \rightarrow t\bar{b} \chi_1^{-}+{\rm h.c.}$, where $\chi_1^0$ is the neutral (wino) LSP  and $\chi_1^{\pm}$ is mainly the charged wino. Thus, gluino pair production leads to events with multiple top quarks and/or $b$-quarks plus missing energy with cross-sections that should be observable with the
2012 LHC data. Furthermore, cascade decays to $\chi_1^{\pm}$ will typically end up as disappearing high $p_T$ charged tracks which could be measured \cite{ATLAS:2012ab, Kane:2012aa}. Neutral wino-like dark matter can also account for the observed PAMELA galactic positron excess  \cite{Adriani:2008zr} as discussed in \cite{Grajek:2008pg,Kane:2009if}. Finally, 
unless there is a sizeable mixing with the neutral higgsino, direct detection of wino dark matter at experiments like Xenon-100 is not possible. In the case of the slightly suppressed higgsino considered in \cite{arXiv:1102.0556} a signal may be observable at next generation direct detection experiments.

Successful electroweak symmetry breaking implies that
the lightest CP-even neutral Higgs boson is light. However, unlike typical TeV-scale phenomenological models, where the precise prediction for the Higgs mass $M_h$ depends on a host of soft parameters, with heavy scalars and suppressed gaugino masses, there is a rather robust and precise prediction for the Higgs mass which essentially only depends on the overall mass scale $m_{3/2}$ and very mildly on $\tan\beta$ for $\tan\beta \gtrsim 5$. For $m_{3/2}$ around 30 TeV, the prediction for the Higgs mass is: $124\,{\rm GeV} \lesssim M_h \lesssim 127$ GeV (except for a suppression
of up to 10 $\%$ at smaller $\tan\,\beta$ which is disfavored) \cite{Kane:2011kj}. For $m_{3/2}$ larger by a factor of two, 
$M_h$ shifts upward by about 2 GeV. Moreover, since in this case one is in the decoupling limit of the MSSM Higgs sector, the Higgs has properties virtually indistinguishable from the SM Higgs. It is quite remarkable that this is precisely the mass where ATLAS and CMS have reported excesses \cite{ATLAS:2012ae}. If these excesses are strengthened with more data, 
this would be the first successful prediction of this set of ideas.

In summary, the suppressed gaugino case predicts the following: 
\begin{itemize}
\item Dark matter has two components: a wino like component and an axion component. The relative fractions of these is not sharply predicted.

\item Gluino pair production is the dominant signal at the LHC with enhanced 
branching fractions to third generation quark final states. 

\item The lightest CP-even Higgs scalar has a mass in the region 122 GeV $\leq M_h \leq$ 129 GeV, with the region 124 GeV $\leq M_h \leq$ 127 GeV being theoretically favored. Furthermore, the couplings of the 
Higgs should be virtually indistinguishable from that of the SM Higgs.

\item No observable signal at Xenon 100, but depending upon the eventual sensitivity,
a signal may be observed at next generation direct-detection experiments.

\item Improvements in sensitivities of precision EDM experiments by few orders of magnitude should also be able to probe the framework. 
\end{itemize}

In the case that gaugino and higgsino masses are not suppressed, they will be of order $m_{3/2}$. 
Then  the only observable signal at the LHC is a Standard Model Higgs boson with a mass of order 125 GeV.
In this case, though, the LSP relic density will be too large. Hence, one has to consider solutions in which the LSP is unstable. An investigation of the breaking of symmetries like $R$-parity in \cite{arXiv:1102.0556} revealed that the effective breaking was too weak to be phenomenologically viable, though there might be other solutions of string/$M$ theory
which enable a `harder' breaking of the symmetry. 

In summary, the un-suppressed gaugino mass case predicts:

\begin{itemize}
\item Dark matter is composed of axions with GUT scale decay constants.

\item The LHC will observe a Standard Model-like Higgs with a mass of ${\cal O}(125)$ GeV. 

\item Similarly, future precision experiments such as the EDM experiments can probe this case also.
\end{itemize}

The key to these generic predictions are the facts above about moduli and axion physics - both of which are of extra dimensional origin. Consideration of moduli and axion physics leads to the mass scale of $m_{3/2} \sim $ 30 TeV, which gives rise to many generic predictions for beyond-the-SM (BSM) physics. Without this string/$M$ theory input, the question - {\it what are the generic predictions of supersymmetry?} - does not have a sharp answer. The rest of this article expands upon all of the above in more detail and can be read independently from this section. 

One result which is included in the remainder of the article, but not mentioned in this summary section is a mechanism for addressing the baryon asymmetry.

\section{String/$M$ theory is a Framework}\label{philo}

String/$M$ theory is a broad {\it framework} for addressing questions in many different physical contexts. For instance,
recently there have been applications to condensed matter systems via the so-called AdS/CMT correspondence. In that sense, string/$M$ theory is no different from quantum field theory (QFT). The QFT framework describes many different physical systems with different numbers of 
spatial dimensions, different symmetry properties, interactions and so on.
The Standard Model of particle physics is a particular example of a QFT whose gauge symmetry happens to be $SU(3) \times SU(2) \times U(1)$.
One does not claim that QFT is not predictive because there are infinitely many examples with other symmetry groups. In this context the question of predictability
of QFT is not even the correct question since we have {\it one} example which describes extremely well the results of all particle physics experiments
conducted thus far. The Standard Model {\it is} a correct description of the physics to the extent that we have tested it and that is enough. 

The different solutions of string/$M$ theory are analogous to the different QFT's and the different solutions can be applied to different physical contexts. The only subset of solutions of string/$M$ theory 
which are of direct relevance for discussing the outcomes of particle physics experiments are those which reduce to the Standard Model in physical contexts where the Standard
Model dominates the physics. In the QFT framework we {\it could} however ask: {\it what are the generic predictions of a QFT with a non-Abelian gauge group, several families of
chiral fermions, hierarchical Yukawa couplings and a mass scale set by spontaneous gauge symmetry breaking\footnote{In particular, here we are thinking of electroweak symmetry breaking.}?} These \emph{generic predictions} would include a rich variety of 3-body decays of the heavier fermions
into the lighter ones spanning many orders of magnitude in lifetime and the existence of massive vector bosons coupling to charged currents. The discovery and measurement of (for example) the $\tau$-lepton, $W$-bosons and their decay properties {\it is} a verification of these generic predictions.

We will analagously describe the broad, ``generic predictions" of solutions to string/$M$ theory which, at energies below the scale of electroweak symmetry breaking reduce to the
Standard Model. Why should we even assume that the string/$M$ theory framework includes the Standard Model? Well, it has been known since the mid eighties
that the {\it generic} solution of heterotic string theory with six compact dimensions is described at low energies by {\it a four dimensional QFT with a non-Abelian gauge symmetry,
several families of chiral fermions and hierarchical Yukawa couplings} \cite{Candelas:1985en} ! Furthermore, analogous statements can be made in other limits of string/$M$ theory; for instance, solutions of $M$ theory in which the seven extra dimensions
form a compact manifold of $G_2$ holonomy with certain kinds of singularity, are also described at low energies by field theories with the same properties \cite{Acharya:2001gy}.Therefore, in solutions of string/$M$ theory which exhibit spontaneous gauge symmetry breaking, a {\it generic prediction} is  the rich variety of 3-body decays, charged currents and massive vector bosons\footnote{A significant fraction of four-dimensional string solutions have Higgs-like fields with symmetry breaking vacua since many solutions
can have fermions with order one Yukawa couplings which radiatively generate symmetry breaking potenials.}. This is a prototypical example of what we mean by a {\it generic prediction}.

The {\it generic predictions} which will interest us in the rest of this review concern physics beyond 
the Standard Model since we would like to confront such predictions with data from the
LHC and results from dark matter experiments and cosmological observations. Spontaneous gauge (for e.g. $SU(2)\times U(1)$) symmetry breaking at a scale much smaller than the Planck scale, is intimately tied to questions such as supersymmetry breaking and the solution to the gauge hierarchy problem. As we will review, understanding how the electroweak scale emerges from string/$M$ theory is also the key to identifying the {\it generic predictions} since
it is closely related to the dynamics of moduli fields and their stabilization. Indeed, much progress on making phenomenological predictions 
has arisen through our improved understanding of moduli physics. This is satisfying,
since it is the moduli fields present in the low energy effective action of string/$M$ theory solutions which sets them apart from QFT's in which the coupling constants are
``constant" and not vacuum expectation values of moduli fields.

Thus, our main assumption is that our Universe is a solution of string/$M$ theory. What we would like to understand is: what are the predictions for physics beyond the
Standard Model that a typical or {\it generic} such solution makes?

Although compactifications to four dimensions 
could give rise to a large set of possibilities, the underlying structure of 
the 10D/11D theory together with its consistency constraints imply that \emph{not} 
all low-energy models can be embedded in a consistent string theoretic framework. Thus, 
there exist `large' classes of theories which are not part of the landscape; rather they are part of a `larger', 
so-called, ``swampland" \cite{Vafa:2005ui}. Further, by making well-motivated theoretical assumptions and including 
all applicable, relevant experimental constraints, it is possible to focus on interesting regions within the full 
landscape of consistent effective field theory models arising from string theory. Now, if one chooses the 
set of experimental constraints which are general and robust in the sense that they apply to low-energy 
models arising from \emph{many} classes of string vacua, 
then the correlated predictions of these classes of models for \emph{other} observables become generic predictions 
of all these classes of models, which can be tested explicitly. We will provide explicit 
examples of the theoretical assumptions, the experimental constraints, as well as the 
correlated predictions for other observables in the following, which will illustrate our approach in detail. 

If any of the experimental results falsify the predictions, this would either rule out those classes of models 
and imply that some of the assumptions have to be relaxed. On the other hand,
 if the results confirm the predictions we will gain an improved 
insight into the structure of the allowed models, leading to more detailed predictions.
 In this sense string theory is completely analogous to quantum field theory.

A low-energy model arising from an underlying string theory has the potential to 
simultaneously provide answers about electroweak symmetry breaking, dark matter, the origin of flavor, matter-antimatter asymmetry 
and connections to other branches of fundamental physics such as early Universe cosmology and quantum gravity. 
Interestingly, while some
generic string theory predictions will hold for all ``weakly-coupled" corners
of string theory (Heterotic, Type IIA/IIB, $M$ theory, $F$-theory etc)
others will only hold for some. That is useful for establishing the general
framework on the one hand, while also understanding which string theories provide the best fit
to the data.

Surprisingly, some important predictions for the LHC and
cosmology can be made without a detailed knowledge of most aspects of the
theory, as we will see! If these predictions are not confirmed, then the implications will be profound.


In section \ref{assumptions}, we clearly set out the broad set of working assumptions under which our claims and arguments are valid. 
In section \ref{stabilize}, we discuss moduli stabilization and 
supersymmetry breaking and the resulting gravitino mass in $M$ theory and other compactifications. In section \ref{mod-spectra}, we provide the outline of a derivation of the upper bound on the mass of the lightest modulus in realistic compactifications as well as some broad consequences for cosmology. From the generic results obtained in this section, 
we will discuss the implications for dark matter in section \ref{dm} and particle phenomenology in section \ref{particlepheno}. 
In section \ref{hep-signals}, the broad consequences for collider physics and precision experiments are discussed. 
This is followed by a brief discussion of the matter-antimatter symmetry within the framework in section \ref{matter-asymm}. 
Finally, we make some comments and provide an outlook for the future in section \ref{conclude}.

\section{Assumptions}\label{assumptions}

In this section we set out clearly the broad set of working assumptions under which our arguments and claims are valid. 
Strictly speaking, none of the assumptions are inevitable consequences of 
string theory compactifications to four dimensions. However, we will explain why each assumption 
is well motivated and sufficiently generic such as to hold true for a large class of solutions. 
For ease of readability, we will also divide the list of assumptions into three categores - 
a) {\it Theoretical}, b) {\it Cosmological}, and c) {\it Model-Building}. 
Finally, although it is hard to order the assumptions in an unbiased manner in terms of 
how strong or how weak they are, some may be more conservative than others.

\vspace{0.3cm}

\emph{{\bf Theoretical}}

\vspace{0.3cm}

The first working assumption is that the vacuum structure of a compactified string theory is determined by an effective 
potential $V_{eff}$ which is a function of all the moduli fields. 
The task is to include all relevant classical and quantum effects in determining $V_{eff}$ and 
then determine its local and/or global minima.  This seems to be a well justified approach in most cases, 
but is difficult to make precise within a theory of quantum gravity (see \cite{Banks:2004xh}). 
However, to the extent that string theory can be considered weakly 
coupled, as far as the vacuum structure and low-scale fluctuations are concerned, it is generally accepted that 
the paradigm of the Wilsonian effective action is applicable, justifying matching to 
field theory just below the string scale, and then following the standard RG paradigm. 
A nice summary of these issues is provided in \cite{Douglas:2006es}.  The philosophy we adopt is that we are only interested in 
string/$M$ theory solutions which could describe our world, 
so we do not need to study the most general set of solutions. 
With this point of view, the Wilsonian effective action paradigm seems quite natural. 

The second assumption we will make is that the solution to the cosmological constant problem is largely decoupled from particle physics considerations. Note that we will still require that the vacuum energy 
vanishes approximately, but assume that additional mechanisms responsible for giving rise to 
the exceptionally tiny value of the cosmological constant have virtually no effect on particle physics. 
This assumption seems to be quite natural and conservative as there is  
no known, measurable, particle physics process in which the \emph{precise} value of the cosmological constant is important. While we cannot be sure until the solution to the cosmological constant problem is agreed 
on, it seems unlikely that knowing its solution will help in calculating the Higgs boson  mass or the relic density of wino-like 
dark matter, etc, and it seems unlikely that not knowing the solution will prevent us from doing such calculations.

Moving on to more ``phenomenological" assumptions, we restrict to compactifications with low scale supersymmetry, as they provide an elegant solution to the hierarchy problem. 
Although low-scale supersymmetry is not known to be a prediction of string theory, it does arise in a variety of different classes of vacua, 
as will be seen in section \ref{stabilize}. In fact, as will also be seen, considerations of axion and moduli physics in a cosmological context leads to
superpartner masses at a scale of tens of TeV. Effectively, therefore we do not consider cases in which superpartner masses are above 100 TeV or so.
The issue of the so-called \emph{little} hierarchy problem i.e. explaining the suppression of the weak scale compared to the supersymmetry breaking scale,
will be discussed in more detail in section \ref{littlehierarchy}. 
In our study, we do not require a fully natural solution to the little hierarchy problem. Of course, if one had a complete underlying theory,  then all of its predictions would be natural by definition.  Issues like ``fine-tuning" only arise when there is an imperfect understanding of the underlying theory.

Treating the apparent unification of couplings in the MSSM as an important clue, we will mainly focus on compactifications with standard grand unification at the Kaluza-Klein scale, 
$M_{KK} \sim M_{GUT} \sim 0.1 M_{st}$, though many of
our results can be extended to other cases. Here $M_{st}$ is the string scale.

Finally, we generally assume that the mass of the lightest modulus field is of order the gravitino mass, $m_{3/2}$ -- a fact which is known to be true for
most explicit string solutions. We will return to this point below and in section
\ref{mod-spectra}.

\vspace{0.3cm}

\emph{{\bf Cosmological}}

\vspace{0.3cm}

We assume that the Hubble parameter in the very early Universe, such as during inflation, is larger than ${\cal O}(m_{3/2})$. 
Since we consider compactifications with low-scale supersymmetry, this implies that the Hubble scale during inflation satisfies 
$H_I \gtrsim 100$ TeV.
Having a large $H_I$ seems to be a natural assumption for the following reason. 
It is known that the slow-roll parameter for simple\footnote{By this we mean single field models of inflation} 
models of inflation $\epsilon \equiv m_{pl}^2\,(\frac{V_I'}{V_I})^2$, where $V_I$ is the inflaton potential 
and $V_I'$ is its derivative with respect to the inflaton, can be written in terms of $H_I$ as:
\be
\epsilon \approx 10^{10}\,\left(\frac{H_I}{m_{pl}}\right)^2
\ee using the value of the primordial density perturbations $\frac{\delta\rho}{\rho} \sim 10^{-5}$. 
The requirement $\epsilon \lesssim 10^{-2}$ for $\sim 60$ e-foldings of inflation to solve the flatness 
and horizon problems implies that $H_I \lesssim 10^{-6}\,m_{pl}$, 
which is the standard fine-tuning in slow-roll inflation models. 
A smaller value of $H_I$ will make $\epsilon$ even smaller, 
implying an even larger fine-tuning than is necessary for inflation. 
Hence, it is natural for $H_I$ to be as large as allowed by data, 
giving rise to $H_I$ (much) larger than ${\cal O}(m_{3/2})$.  
Another assumption we make which is relevant for cosmology is that not 
\emph{all} the moduli are stabilized close to an enhanced symmetry point: 
it is clear that in a generic case this will be satisfied, 
only under extremely special circumstances could it be violated, if at all. 

It is worth addressing the caveats that go along with the above ``cosmological" assumptions and possible alternatives. For example,  it has been argued that string compactifications with 
stabilized moduli typically lead to the requirement that $H_I \lesssim m_{3/2}$ since otherwise the purported stabilized vacuum tends 
to get destabilized from the large positive contributions to the vacuum energy during inflation \cite{Kallosh:2004yh}. 
However, this requirement is naturally relaxed if the location 
in field space at which inflation ends is far apart from the location of the true late-time minimum of the potential.  
Then, after the end of inflation the various moduli fields are far away from their late-time minima and evolve towards it. 
If there are sufficient ``friction terms" present in the evolution in the form of matter or radiation or other moduli fields which are necessarily present, 
then the fields can relax to their true minima without destabilizing the potential as argued in \cite{Conlon:2008cj}. 
One could try to construct low scale inflation models with $H_I \lesssim m_{3/2}$, 
but it is not clear how generic or well-motivated they are within a string-theoretic framework, especially with $m_{3/2}$ 
around the 10 TeV scale. Similarly, one could try to stabilize moduli with masses much larger than $m_{3/2}$. 
Although this may be possible with fine-tuning in compactifications with a single modulus \cite{Kallosh:2006dv}, it is 
\emph{extremely} non-generic and does not work for 
realistic compactifications with many moduli, see section \ref{mod-spectra} for details. 
So, these possibilities will not be considered. 

Finally, it is important to understand the relevant issues in models in which the scale of supersymmetry breaking is quite low. 
In this case the gravitino mass $m_{3/2}$ is much smaller than the TeV scale. This implies that the lightest modulus mass is of ${\cal O}(m_{3/2}) \ll $ TeV. 
This would give rise to a very serious moduli problem since such moduli still dominate the energy density of the Universe, but decay long after the BBN era. 
Constructing string-theory models of inflation with $H_I \lesssim m_{3/2}$ with such small values of $m_{3/2}$ in  
order to avoid the problem appears even less natural than that for cases with $m_{3/2} \sim $ TeV. 
Alternatively, one could try to appeal to models with thermal inflation \cite{Lyth:1995ka} 
which produces sufficient late-time entropy to dilute the abundance of the oscillating moduli. 
However, such models have rather special requirements which appear 
quite \emph{ad hoc} from a top-down point of view. The authors of \cite{Fan:2011ua} tried to provide a well-motivated model of thermal inflation which 
could also provide a solution to the strong-CP problem. In this class of models, the QCD axion solving the 
strong CP problem arises from low-energy field theory and does not originate from a string compactification. 
Since the QCD axion must have a very flat potential to solve strong-CP, it is natural for the axion partner, 
the saxion, to also have a sufficiently flat potential giving rise to thermal inflation. 
However, there are many challenges in constructing viable models. 
From the theoretical point of view, it is quite challenging to construct a global 
symmetry which is preserved at the level required to solve the strong CP problem. Phenomenologically, 
it is challenging to construct models which satisfy constraints from the observed $\gamma$ ray background \cite{Fan:2011ua}.

From a top-down point of view, there does not exist a strong motivation for a low-energy field-theory solution to the strong CP-problem. 
Within string theory, it is much more natural for the QCD axion to arise from string theory as string theory naturally provides us with many axions. 
As will be discussed in section \ref{axion-dm}, there exist natural moduli and axion stabilization mechanisms which stabilize all moduli and axions in such a way that 
the moduli (including the QCD saxion) receive masses of ${\cal O}(m_{3/2})$ while the axions receive masses exponentially suppressed relative to $m_{3/2}$. 
One of these exceptionally light axions can then naturally serve as the QCD axion, providing a solution to the strong-CP problem within string theory 
\cite{Acharya:2010zx}. In light of such a mechanism, there exists little motivation for QCD saxion thermal inflation, if such a mechanism is possible at all.

\vspace{0.3cm}

\emph{{\bf Model-Building}}

\vspace{0.3cm}

We finally turn to assumptions about model-building. There has been considerable progress in 
string phenomenology in this regard, and large classes of string models have been constructed with quasi-realistic gauge groups and matter content. 
The origin of flavor has also been addressed in several different classes of solutions \cite{Heckman:2008qa, Leontaris:2010zd}. 
For our purposes, therefore, we will make the reasonable assumption that the 
string/$M$ theory compactification is such that, at low energies it gives rise to
SM gauge group with the matter content of the MSSM. 
The precise unification of gauge couplings, as well as successful radiative electroweak symmetry breaking in the MSSM 
provide strong support for such an assumption. Moreover, explicit string solutions with \emph{precisely} the MSSM content have been constructed \cite{Braun:2005nv}. 

We assume that the visible sector is weakly coupled, i.e. all the SM fermions 
as well as the Higgs fields are elementary (as opposed to composite) so that the standard Higgs mechanism gives rise to electroweak 
symmetry breaking in the effective low-scale theory.  As far as supersymmetry breaking is concerned, because of the moduli problems
discussed above, we assume gravity mediation with a ``hidden sector" of supersymmetry breaking.
This is generic in eleven dimensional $M$ theory compactifications in which the extra dimensions form a $G_2$-manifold.
There, non-Abelian gauge fields are localized along three-dimensional submanifolds of the seven extra dimensions. In
seven dimensions, two three-manifolds generically don't intersect, there are no matter fields charged under both the Standard Model
and hidden sector gauge symmetries. 
For other compactifications, this is more model-dependent but can be satisfied; in any case, generically it must be satisfied in order
to avoid the moduli problem.

Finally, we assume that there are no other flavor-universal  $R$ or non-$R$ global symmetries (such as a $PQ$ symmetry) 
at low energies. This is a natural assumption since global symmetries are generically broken in the presence of gravity by Planck suppressed operators, and within gravity mediation these Planck suppressed operators are relevant. Also, the vanishingly tiny value of the cosmological constant implies that the superpotential does not vanish in the vacuum obtained after moduli stabilization, which explicitly breaks any potential $R$-symmetry.  However, we do assume that there is at least an approximate (it may be exact) discrete symmetry which keeps the LSP sufficiently long-lived, and hence provides a WIMP DM candidate\footnote{This is for the case with suppressed gaugino masses, for unsuppressed gaugino masses to be viable the $R$-parity must be sufficiently broken.}.  Again, such symmetries naturally arise within string/$M$ theory solutions with grand unification. 

It is worth mentioning that although we make some model-building assumptions, the crucial results 
that the lightest modulus mass is ${\cal O}(m_{3/2})$, and therefore $M_{scalars} \gtrsim 30$ TeV, are derived from the 
underlying theory by just imposing cosmological constraints, and are therefore not model-dependent.

\vspace{0.3cm}

\section{ Moduli Stabilization and Susy Breaking}\label{stabilize}

Based on the philosophy and assumptions outlined in the previous section, we summarize the results about moduli stabilization and 
supersymmetry breaking which will be relevant for low energy particle physics. We are interested in compactifications of string/$M$ theory to 
four dimensions which preserve  ${\cal N}=1$ supersymmetry. In the limit in which the string coupling is small and the extra dimensions are small (but large enough that
the supergravity approximation is valid), 
the low energy four-dimensional theory obtained in ${\cal N}=1$ compactifications of \emph{all} corners of string/$M$ theory is ${\cal N}=1,\,D=4$ supergravity.
Here ``low energy" 
refers to energies far below the compactification scale, or Kaluza-Klein (KK) scale. As explained earlier, we will consider cases where $M_{KK} \sim M_{GUT}$. Since $M_{KK}$ is only 
determined \emph{after} moduli stabilization, the above condition has to be checked self-consistently.

${\cal N}=1,\,D=4$ supergravity is completely specified at the two-derivative level by three functions\footnote{See \cite{Nilles:1983ge} for a review. Strictly speaking, the lagrangian depends upon two functions, but it is convenient to use three. }  :

\begin{itemize}

\item  The superpotential $W$, which is a holomorphic function of the chiral superfields. 
$W$ is not renormalized in perturbation theory but receives non-perturbative corrections in general.
\item  The gauge kinetic functions $f_a$ for each gauge group $G_a$, which are also holomorphic functions of the chiral superfields.
\item The K\"{a}hler potential $K$, which is a real non-holomorphic function of the chiral superfields and their complex conjugates. 
Unlike $W$ and $F$, $K$ receives corrections to all orders in perturbation theory. The finiteness of the 
string scale gives rise to corrections in powers of $\left(\frac{l_s}{V^{1/6}}\right)$ 
where $l_s$ is the string length and $V$ is the (stabilized) volume of the extra dimensions
\footnote{This can be easily generalized to the 11D $M$ theory case.}.
For values of $V$ which correspond to the unification scale $M_{GUT}$, 
these corrections are small, as in the ones discussed below. 
\end{itemize}
Different compactifications of string/$M$ theory give rise to different functional forms for $K$, $W$ and $f$ in general, 
although we will see later that phenomenologically realistic solutions arising from different corners of string theory share many common features. 

The fields in the four dimensional theory include the moduli, axions and charged matter fields (both visible and hidden) as well as their superpartners. 
However, since the moduli (and some hidden sector matter) fields generically acquire large {\it vevs} ($\sim M_{st}$) while the visible matter fields 
must have vanishing {\it vevs}\footnote{The Higgs {\it vev}, although non-zero, 
is much smaller than the typical moduli {\it vevs}.}, it is a good approximation 
to first study the moduli and hidden matter potential, susequently adding visible sector matter fields 
as an expansion around the origin of field space. It is important to also make sure that effects 
which could induce \emph{vevs} for SM-charged matter fields are not present \cite{Blumenhagen:2007sm}.

The gravitino mass $m_{3/2}$ is given in ${\cal N}=1$ supergravity by:
\ba\label{m32}
m_{3/2}  = e^{K/2}\,\frac{\langle W \rangle}{m_{pl}^2} = \frac{\sqrt{\sum_i\,\langle F^i F_i\rangle}}{\sqrt{3}\,m_{pl}}
\ea
where $F_i$ are the $F$-terms (defined as derivatives of $K$ and $W$ wrt to the scalar fields: $F_i=e^{K/2}(\partial_i\,W+\partial_i\,K\,W)$); a non-zero
expectation value for any of the $F_i$ implies supersymmetry is broken. $m_{pl}$ is the 
reduced Planck scale $m_{pl}\equiv M_{pl}/\sqrt{8\pi}$, and in the second equality we have used the fact that the 
cosmological constant is vanishingly small. Hence in such vacua the gravitino mass is the order-parameter of supersymmetry breaking. 
Now, since we are interested in classes of vacua with low-energy supersymmetry to provide a solution to the hierarchy problem, 
$m_{3/2}$ must be much smaller than $m_{pl}$. This implies\footnote{A large and negative $K$ corresponds to a Kaluza-Klein scale much
less than $M_{GUT}$} that $\langle W\rangle$ or $\bar{F}\equiv\sqrt{\sum_i\,\langle F^i F_i\rangle}$ 
has to be much suppressed relative to $m_{pl}$. In order to discuss these dynamical issues, 
it will be useful to illustrate the results with an example. We will then see how the results generalize to other compactifications as well.

\subsection{$M$ theory compactifications on $G_2$ manifolds}\label{mtheory}

Compactifications of 11D $M$ theory to four dimensions preserve ${\cal N}=1$ supersymmetry if the metric on the seven extra dimensions has holonomy group equal to the exceptional Lie group $G_2$ \cite{Papadopoulos:1995da}.

Phenomenologically relevant compactifications with non-abelian gauge symmetry and chiral fermions can 
only arise from $G_2$ manifolds endowed with special kinds of singularities.
In particular, non-abelian gauge fields are localized along three-dimensional submanifolds 
inside the internal space \cite{Acharya:1998pm} while chiral fermions are supported at points in the extra dimensions where there are conical singularities 
of particular kinds \cite{Acharya:2001gy}. The gauge fields and chiral fermions of course also propagate in the four large space-time dimensions. 
Although many examples of smooth $G_2$ manifolds have been constructed \cite{math}, an explicit construction of {\it compact} $G_2$ manifolds with all the
singularities required to give rise to phenomenologically relevant solutions has proven so far to be too challenging technically, though many such manifolds are strongly
believed to exist. ``String dualities" imply the existence of many examples:
for instance, the duality between $M$ theory and heterotic string and Type IIA compactifications.  
We will thus assume that singular $G_2$ manifolds supporting non-abelian gauge theories and chiral fermions exist and use the fact that 
enough is known about the $K$'s, $W$'s and $f$'s which arise from $G_2$-manifolds in order to  proceed.
Many properties of the four-dimensional ${\cal N}=1$ theory relevant for particle physics can be derived from a Kaluza-Klein reduction of 11D supergravity to four dimensions, 
which is the low energy limit of $M$ theory. 

At low energies, $M$ theory is described by eleven dimensional supergravity theory which contains a metric, a 3-form gauge field ($C$) and a gravitino.
We will not be interested in solutions in which there is a non-trivial flux for the field strength of $C$ along the extra dimensions:
although fluxes can stabilize moduli   
\cite{Acharya:2002kv}, they do not generate a hierarchy between the Planck scale and the gravitino mass. 
We will see in section \ref{other} that unlike $M$ theory, 
fluxes do play an important role in Type IIB string theory. 

In $M$ theory compactifications on a $G_2$-manifold, all the moduli fields are geometric - 
they arise as massless fluctuations of the metric of the extra dimensions \cite{Acharya:2004qe}. Since these moduli $s_j$ are {\it real} scalar fields\footnote{The
subscripts $i,j,k,..$ are used to enumerate the moduli fields and their superpartners.}, in order to
reside in the {\it complex} chiral supermultiplets required by supersymmetry, additional real scalar fields must also be present. These additional fields are
the axions $a_j$ which arise as the harmonic fluctuations of $C$ along the $G_2$-manifold. The moduli and axions pair up to form {\it complex} scalar fields
which are the lowest components of chiral superfields $\Phi_i$ in the effective 4d supergravity theory:
\be\label{defn-M}
\Phi_j = a_j + i s_j + fermion \;\; terms
\ee

\subsubsection{ Moduli and Scales in $M$ theory on a $G_2$-manifold}\label{scales-M}

For future reference this subsection summarizes the relations between the $G_2$- moduli, the volume of the extra dimensions and the gauge couplings. More precise relations are given in \cite{Friedmann:2002ty}.
The volume $V_7$ of a $G_2$-manifold is a homogeneous function of the moduli of degree $7/3$. For instance, if the volume of a $G_2$ manifold
is dominated by a single modulus field, then $V_7 \sim s^{7/3}$. Roughly speaking, {\it it is useful to think of the moduli vevs as parametrizing
the volumes of a set of independent three-dimensional submanifolds} of the $G_2$-manifold, in units of the 11d Planck length. 
So, if $V_7$ is dominated by a single term, we think of the volume of the $G_2$-manifold as being dominated by a single three-cycle $Q$ with volume
$Vol(Q) \sim s$. Non-Abelian gauge fields are localized along three-dimensional submanifolds, hence the effective gauge coupling $g^2_{YM}$ is related
to the volume of the three-manifold as ${4\pi \over g^2_{YM}} = 1/\alpha = Vol(Q)$. Consider then a $G_2$-manifold such that the three-manifold which supports
the Standard Model (unified) gauge group dominates $V_7$. We then have that
\be
V_7 \sim {1 \over \alpha_{GUT}^{7/3}} \sim (25)^{7/3}
\ee
where we use the fact that these volumes are understood to be given at the GUT scale and that $\alpha_{GUT} \sim 1/25$ with the MSSM field content. 
We then further infer that
\be
m_{pl}^2 \sim V_7 M_{11}^2 = {M_{11}^2 \over \alpha_{GUT}^{7/3}}
\ee

and, because the volume of $Q$ is given by $1/\alpha_{GUT}$ that

\be
M_{KK} \sim M_{GUT} \sim {M_{11} \alpha_{GUT}^{1/3}}
\ee

Thus, a value of $\alpha_{GUT} \sim 1/25$ gives a set of relations consistent with Newtons constant, $M_{GUT} \sim 2 \times 10^{16}$GeV and
$M_{11} > M_{GUT}$. The latter fact is required for validity of the low energy effective field theory approximation. 

More generally if we assume that the $G_2$-manifold is more or less isotropic then we expect the vevs of all moduli to be of the same order, and
hence the above scalings with $\alpha_{GUT}$ will still hold true. We now return to discuss the potential for the moduli in more detail.

\subsubsection{Hierarchies are Generic in $M$ theory}\label{hierarchy-M}

All the moduli $\Phi_j$ of $M$ theory are invariant under shift symmetries \cite{Acharya:2004qe}:
\be\label{shift}
\Phi_j \rightarrow \Phi_j + c_j
\ee
with $c_j$ being an arbitrary constant.The origin of the shift symmetries can be understood as follows.  The real parts of the moduli $\Phi_j$, denoted by $a_j$ in (\ref{defn-M}), arise from the Kaluza-Klein (KK) reduction of the three-form (antisymmetric tensor field with three indices) in eleven dimensions to four dimensions. The underlying gauge symmetry of this three-form in higher dimensions reduces to shift symmetries for the individual axions in four dimensions: $a_j \rightarrow a_j + c_j$. With ${\cal N}=1$ supersymmetry, the $a_j$ combine with the modes arising from the KK reduction of the metric in eleven dimensions to form chiral superfields whose scalar components are $\Phi_j$.

The above symmetries imply that the effective superpotential in four dimensions, $W$, which must be a holomorphic function of the $\Phi_j$, does not contain `perturbative' terms (i.e. terms polynomial in the $\Phi$'s). Hence 
{\it the perturbative superpotential for the moduli vanishes exactly}. This is a key point which distinguishes $M$ theory on a $G_2$-manifold from other compactifications
such as Type IIB and heterotic string theories on a Calabi-Yau manifold. For instance, Calabi-Yau manifolds generically have complex structure moduli; since these moduli
are already complex fields, the corresponding supermultiplets do not have a shift symmetry and, consequently, the superpotential can contain perturbative contributions
dependent on these fields.
 
However, since axionic shift symmetries are generically broken by non-perturbative effects the superpotential will not be zero in general. For instance, if there is an asymptotically
free gauge interaction present then the corresponding strong gauge dynamics at low energies will {\it necessarily} generate a non-perturbative superpotential proportional to
(the cube of) the dynamically generated strong coupling scale ($\Lambda$): $W \sim \Lambda^3 \sim e^{\frac{-b}{\alpha_{Q}}} m_{pl}^3$ 
(in this case $1/b$ is the one loop $\beta$-function coefficient
of the hidden sector gauge theory and $\alpha_Q$ is its fine-structure constant).  
More generally, `pure' membrane instantons can generate terms in the superpotential \cite{Harvey:1999as}. In fact,
{\it every term in the superpotential can be associated with a 3-cycle and will be proportional to $e^{i b N^j \Phi_j}$}. Here, the $N^j$ are the $b_3(X)$ integers specifying
the homology of the 3-cycle and $b$ is a number characterising the given instanton contribution. Obviously, different instanton contributions will have 
different values for $b$ and $N_j$.

For solutions
of $M$ theory for which the 4d supergravity approximation is valid, the KK scale is below the 11d Planck scale $M_{11}$ and all of these non-perturbative contributions,
which are of order $e^{-b{M_{11}^3 \over M_{KK}^3}}$, are {\it exponentially small}. Thus, on general grounds, one expects $M$ theory compactifications without
flux to generate a very small expectation value for $W$, which in turn implies an exponential {\it hierarchy} between $m_{3/2}$ and the Planck scale $m_{pl}$. 
Thus, we see that $M$ theory on a $G_2$-manifold without flux is an ideal framework for addressing the hierarchy problem.
The key questions then
become: a) can the moduli potential generated by strong hidden sector gauge dynamics also stabilize the moduli?  b) does this potential spontaneously break supersymmetry? These questions
were answered affirmatively in \cite{hep-th/0606262,hep-th/0701034}, thereby providing a proof of the `lore' that hidden sector strong dynamics could i) generate the hierarchy between the
Planck and weak scale, ii) stabilize the moduli fields and iii) spontaneously break supersymmetry.

It is further important to note that, in a region of field 
space where the supergravity approximation
is valid, volumes are larger than one in 11d units, so the contributions to the potential from strong gauge dynamics is exponentially larger than the purely 
membrane instanton effects; this is because in the former
case $b$ is proportional to ${1 \over d}$ where $d$ is a one-loop beta-function coefficient -- typically an integer larger than unity -- whereas such a `suppression' of $b$ is not
present for membrane instantons. Hence, if strong gauge dynamics is present at low energies it will dominate the moduli potential.

We will now review the results of \cite{hep-th/0606262,hep-th/0701034, arXiv:0810.3285} in more detail. The simplest possibility is to consider a $G_2$-manifold with a single hidden sector
interaction which becomes strongly coupled at some scale much smaller than $M_{KK}$. For instance, this could be given by $SU(N)$  super Yang-Mills theory with no light
charged matter. 
In terms of the supergravity quantities, 
the inverse gauge coupling $\alpha_h^{-1}$ can be identified with the imaginary part of the gauge kinetic function for the hidden sector gauge theory ($f_h$), while 
the strong coupling scale $\Lambda$ can be identified with the non-perturbative superpotential alluded to above \cite{Acharya:2004qe}: 
\be W \sim \Lambda^3 =  e^{\frac{2\pi i}{N}\,f_h}\,m_{pl}^3.\ee  In $M$ theory, the K\"{a}hler potential $K=-3\,\log(V_7)$ \cite{Beasley:2002db}. Substituting $K, W, f_h$ into the
formula for the supergravity potential and minimising indeed shows {\it formally} that all the moduli can be stabilized in this case. However, the vacuum is located in a region where
the supergravity approximation is not valid since the the 3-cycle volume is negative.

Following this, we  considered two hidden sector gauge theories -- both super Yang-Mills without light charged matter. There are thus two dominant terms in the superpotential, 
characterized by two integers $P$ and $Q$, the one-loop $\beta$-function coefficients.
\be
W = A_1  e^{\frac{2\pi i}{P}\,f_{h1}}  + A_2  e^{\frac{2\pi i}{Q}\,f_{h2}}
\ee where we introduced two {\it constant} normalizations $A_1$ and $A_2$ and set $m_{pl}=1$. The normalizations are constant in $M$ theory due to the axionic shift symmetries.
Further simplification arises by assuming that $f_{h1}$ and $f_{h2}$ are proportional to one another, though the more general case was analyzed in \cite{Acharya:2011kz}. We will describe the
results of the simplified cases here for ease of exposition, thus we set $f_{h1}=f_{h2}=f=\sum_{i=1}^{N}\,N^i\,\Phi_i$, since the proportionality constant can be absorbed into re-defining $Q$. Here $i$ runs over all the $N$ moduli in general and $N^i$ are positive integers.

With two strong hidden sector interactions the supergravity potential has many stable vacua in which all the moduli are stabilized.
The fact that there are many vacua should 
not come as a surprise since a sufficiently generic potential for $N$ fields will possess of order $2^N$ critical points. 
Moreover many of these vacua are
in regions where the supergravity approximation is applicable and in these vacua the hidden sector coupling $\alpha_h << 1$. 
One can (semi-analytically) study the potential
close to the minima in an expansion in $\alpha_h$. Let us consider the supersymmetric (anti-de Sitter) vacuum\footnote{The formulae for the non-supersymmetric vacua are
very similar}.

Here, one finds
\be
\frac{1}{\alpha_h} = {1 \over 2\pi} {PQ \over Q-P}\,\log\left({A_1 Q \over A_2 P}\right)
\ee
 and the (dimensionless) moduli vevs are fixed to be of order
\be
s_i \sim {1 \over N_i} {1 \over \alpha_h};\;\;i=1,2,..,N
\ee 
As mentioned above, one can think of these moduli vevs as volumes (in eleven dimensional Planck units) of various three-manifolds inside the seven-dimensional $G_2$ manifold. 
We can now calculate the gravitino mass as a function of $P,Q$ and $\alpha_h$ up to numerical constants:
\be
\frac{m_{3/2}}{m_{pl}} = e^{K/2}\,\frac{W}{m_{pl}^3}\sim A_2 \frac{e^{\frac{2\pi i}{c}\,f_h}}{V_7^{3/2}} 
=A_2 {|Q-P| \over Q} \alpha_h^{7/2} e^{-\frac{2\pi}{Q \alpha_h}}
\ee
so that a value of $\alpha_h \sim 1/25$ with $A_2=1, Q=8, P=7$ for example, gives $m_{3/2} \sim 100$ TeV.
Note that $\alpha_h \sim 1/25$ can arise from values of $P$ and $Q$ $\leq 10$ and normalization
constants such that the logarithm in the formula for $\alpha_h$ is $\geq 1$.

Another requirement for a realistic compactification is a de Sitter (dS) vacuum i.e. positive cosmological constant. 
It is well known that de Sitter vacua do not arise in the classical limit of string/$M$ theory \cite{Maldacena:2000mw}.
A review of de Sitter space in string/$M$ theory is given in \cite{Rummel:2011cd}.
One can interpret this result as ``the classical potential for moduli fields does not possess de Sitter vacua". In the examples studied above, though the
potential is generated through quantum effects, it is exponentially close to the classical limit in the sense that all terms in the potential are exponentially small.
Hence, it is not surprising that all of the vacua found had negative cosmological constant. Therefore, we expect that in a would be de Sitter vacuum that the vacuum
energy is dominated by a field which is not a modulus of the $G_2$-manifold. 
In \cite{hep-th/0701034,arXiv:0810.3285}, it was shown 
that including matter fields charged under the hidden sector gauge symmetries leads straightforwardly to a vacuum with a positive cosmological constant in which the 
dominant contribution to the vacuum energy arises from the $F$-term of a hidden matter field.
This turns out to be quite relevant for many phenomenological features, as will be seen below. 

Finally, it is important to mention the stabilization of axions, which are the imaginary parts of the 
complex moduli fields.  The moduli stabilization mechanism stabilizes \emph{all} the moduli but gives a mass of ${\cal O}(m_{3/2})$ to only one combination of axions. 
The masses of the other axions are generated by higher order instanton effects which make them exponentially suppressed relative to $m_{3/2}$ \cite{Acharya:2010zx}. This is crucial for a solution to the strong CP-problem discussed in section \ref{axion-dm}.

\subsection{Type IIB and Other Compactifications}\label{other}

Here we consider moduli stabilization and supersymmetry breaking in other branches of string theory. 
We will discuss Type IIB compactifications here as progress towards phenomenologically viable moduli stabilization was first made for these compactifications 
\cite{Dasgupta:1999ss, Giddings:2001yu, Kachru:2003aw}, stimulating a lot of activity \cite{Denef:2005mm,Balasubramanian:2005zx, Blumenhagen:2008kq,Cicoli:2011qg}. 
The Type IIB compactifications are also better understood from a technical point of view and compactification manifolds with the required properties to stabilize
 moduli can be explicitly constructed. We will also see that the moduli stabilization mechanism described above in the $M$ theory case can be essentially carried over to the Type IIB
 case with minimal differences. Since it is possible to construct explicit compactifications in Type IIB satisfying the criteria for moduli stabilization, this proves the 
robustness of the physical ideas which are crucial in stabilizing all moduli in $M$ theory and Type IIB compactifications. At the end, we will briefly comment on Type IIA 
and Heterotic compactifications.

In Type IIB compactifications to four dimensions, non-abelian gauge theories can arise on the worldvolumes of D-branes - such as D7-branes wrapping a four-dimensional manifold 
inside the six-dimensional internal manifold. Chiral fermions arise from open strings at the intersection of two D7-branes \cite{Berkooz:1996km}. There are three different kinds 
of moduli in these compactifications - complex structure, dilaton and K\"{a}hler moduli. Unlike $M$ theory compactifications where all moduli were invariant under a shift symmetry,
 in this case only the K\"{a}hler moduli are invariant. Therefore, the perturbative superpotential can depend upon the complex structure and dilaton.
 It is possible to stabilize these moduli supersymmetrically at a high scale ($\sim M_{KK}$)  by an appropriate choice of fluxes \cite{Kachru:2003aw, Douglas:2006es}. However,
 the K\"{a}hler moduli are not stabilized by this mechanism. 
Non-perturbative effects can stabilize the K\"{a}hler moduli just as in the $M$ theory case. Therefore, these moduli are generically
much less massive 
than the complex structure moduli. 
It is convenient to first integrate out the heavier moduli, which gives a constant contribution to the superpotential - $W_0$. This has to be combined with
 the non-perturbative contributions to stabilize the K\"{a}hler moduli. As explained earlier, in order to solve the Hierarchy problem the value of the superpotential in the vacuum must
 be much smaller than $m_{pl}^3$ (if one does not want the extra dimensions to be extremely large). Hence, $W_0$ must be very small (or zero). This can be arranged by a proper choice of fluxes in Type IIB, but involves some tuning 
\cite{Douglas:2004qg, Dine:2004is}.  Note that in the $M$ theory case $W_0=0$ precisely, so the entire superpotential is non-perturbative naturally.

Vacua also exist in Type IIB theory with $W_0={\cal O}(1)$. In these vacua, if one includes the leading perturbative corrections to the K\"{a}hler potential, the moduli are
stabilized at large values in which the volume of the Calabi-Yau manifold is exponentially large. This is the so-called LARGE Volume Scenario (LVS) and was developed in \cite{Balasubramanian:2005zx, Conlon:2005ki}. LVS vacua exist partly because of a balancing between the perturbative and non-perturbative contributions to the potential which
give rise to an exponentially large volume for the extra dimensions or, equivalently, an intermediate string scale.
One obtains a hierarchy between $m_{3/2}$ and $m_{pl}$ precisely because of the exponentially large volume -- which corresponds
to a large and negative expectation value for the K\"{a}hler potential.
A variety of different possible phenomenological scenarios are possible in the LVS scheme resulting in
different mass hierarchies between supersymmetric particles. 
The existence of LVS vacua is also closely tied to the fact that, in the classical limit, the low energy effective supergravity
theory describing Type IIB compactifications exhibits what is called ``no-scale structure". This implies, among other things, 
that the vacua of the classical potential have zero vacuum energy and is the reason why the perturbative corrections have such a significant effect when $W_0$ is not tuned to be small. With a lower string scale, LVS vacua do not generically give rise to grand unification at around 10$^{16}$GeV. For this and related reasons, some of the generic predictions we make may not always apply to LVS solutions.   

A generic  Type IIB compactification has many K\"{a}hler moduli in general, but most of the moduli stabilization mechanisms in many explicit examples work only for a 
few K\"{a}hler moduli. There is however one robust mechanism, valid for small $W_0$, which stabilizes  
\emph{all} K\"{a}hler moduli in a compactification with many K\"{a}hler moduli 
with minimal ingredients, and is inspired by results obtained in $M$ theory \footnote{Another possible approach to stabilize more than one K\"{a}hler moduli in an LVS-like scenario is to use a diagonal del-Pezzo divisor to stabilize the overall volume. The remaining K\"{a}hler moduli are stabilized by a combination of $D$-term constraints and string loop corrections \cite{Cicoli:2011qg}.}. It was shown in \cite{arXiv:1003.1982} that if the non-perturbative superpotential depends on a linear 
combination of all K\"{a}hler moduli, it is possible to stabilize all K\"{a}hler moduli as long as the four-manifold supporting the instanton or gaugino condensate satisfies a certain 
mathematical property, namely it is a ``rigid ample divisor".   Thus, the qualitative result of stabilizing all moduli with low-scale supersymmetry can be obtained 
in this class of compactifications as well. Explicit Calabi-Yau manifolds satisfying the above criteria were constructed in \cite{arXiv:1003.1982}. These compactifications share the 
interesting feature with the $M$ theory case that all but one axion are stabilized with exponentially suppressed masses relative to $m_{3/2}$, which is crucial for solving the strong-CP 
problem. Finally, in order to generate a vacuum with positive cosmological constant, vacuum energy contributions from non-moduli sources must be included, as in $M$ theory. In Type IIB 
compactifications, in addition to possible $F$-term contribution arising from a hidden matter sector as in the $M$ theory \cite{Lebedev:2006qq}, there could be contributions arising from 
$D$-terms \cite{Burgess:2003ic} or from explicit supersymmetry-breaking effects as well \cite{Kachru:2003aw}.
However, many consequences for phenomenology do not depend on the details as long as certain simple conditions are satsified, as we will explain in the following subsection.

Finally, let us briefly comment on moduli stabilization in Type IIA and Heterotic compactifications. In these cases, fluxes can stabilize some moduli 
\cite{DeWolfe:2005uu, Acharya:2006ne, Gukov:2003cy} but generically fail to generate the hierarchy. However, a better understanding of these compactifications may eventually lead to
progress in demonstrating the existence of vacua with low energy supersymmetry in particular classes, see for example \cite{Anderson:2011cza,Dundee:2010sb}.

\section{Moduli Spectra and Implications}\label{mod-spectra}

In this section, we provide some insight into the spectrum of the moduli in general, and the lightest modulus mass in particular, relative to the gravitino mass. We first explain a general result about the lightest modulus mass which works in all cases satisfying the supergravity approximation and is independent of the details of moduli stabilization. This provides a significant bound on the lightest modulus mass  ${\cal M}_{min}={\cal O}(1)\,m_{3/2}$ in cases in which all scales are set by one scale (such as $m_{pl}$), but not when there are other scales $\Lambda$ ($\ll m_{pl}$) present as well.  However, we then show that in realistic frameworks in which the moduli are stabilized by the mechanisms considered in section \ref{hierarchy-M} and \ref{other}, it is possible to derive the same bound even in cases with additional scales $\Lambda \ll m_{pl}$. We will then discuss the the range of the gravitino masses in viable compactifications, and briefly describe the implications for cosmological history.

\subsection{Lightest Modulus: General Supergravity Result}\label{gen-result}

The general supergravity result was derived in \cite{Acharya:2010af}, building on the work of \cite{Denef:2004cf}. The basic argument is as follows. One considers the mass-matrix ${\cal M}^2$ for all the scalar fields in the true dS vacuum with broken supersymmetry, which is positive definite by assumption. Then, one can use the theorem that its smallest eigenvalue ${\cal M}_{min}^2$ is smaller than $\xi^{\dag}\,{\cal M}^2 \xi$ for any unit vector $\xi$. Then, choosing a direction in scalar field space which corresponds to that of the  sGoldstino (the superpartner of the Goldstino), one can show that: 
\ba
{\cal M}_{min}^2 = m_{3/2}^2\,\left(2+\frac{|r|}{m_{pl}^2}\right) 
\ea where $r$ is the ``holomorphic sectional curvature" in the space of scalar fields \cite{Denef:2004cf}, evaluated in the sGoldstino directions. Now, if the only scales in the problem are set by $m_{pl}$, then $\frac{r}{m_{pl}^2}={\cal O}(1)$. For example, this is the case in $M$ theory compactifications where all the moduli arise from the metric and the only scale is set by the 11D planck scale $M_{11}$, which determines both $m_{pl}$ and $m_{3/2}$ in terms of dimensionless constants after moduli stabilization. This gives the result: \ba {\cal M}_{min} = {\cal O}(1)\,m_{3/2} \label{mmod}\ea which we set out to prove. 

In other string compactifications, however, there are different kinds of moduli, such as the dilaton, complex structure and K\"{a}hler moduli. We focus on the Type IIB case for concreteness. In this case, the K\"{a}hler moduli are similar to the moduli in $M$ theory, but it is possible in general that additional scales $\Lambda \ll m_{pl}$  may be present for the sGoldstino, due to the existence of other kinds of moduli. In this case, $|r|$ is enhanced by the ratio $\frac{m_{pl}^2}{\Lambda^2}$, so that $\frac{|r|}{m_{pl}^2} = {\cal O}(1)\,(\frac{m_{pl}^2}{\Lambda^2})$ \cite{Acharya:2010af}. In these cases, the general supergravity result, although correct, does not provide a useful bound.

\subsection{Result with Stabilized Moduli }\label{stabilize-result}

We now outline the argument that in realistic cases where all the moduli are stabilized by the mechanism explained in sections \ref{hierarchy-M} and \ref{other}, the result (\ref{mmod}) holds even in the presence of additional scales. A detailed derivation will appear shortly \cite{Kumar:2012xx}.

One is interested in the following superpotential and K\"{a}hler potential:
\ba\label{WK}
W &=& W_0 + A\,e^{-b\,T_D};\;\;T_D=\sum_i\,{n_i\,T_i}\nonumber\\
K &=& - \gamma \log{(V_X)}
\ea
in units of $m_{pl}$. Here $V_X$ is the volume of the internal manifold in string or 11D units, $A, b, \gamma$ are numerical coefficients, and $n_i$ are positive integers. This form is natural in Type IIB compactifications, but it can be easily generalized to the $M$ theory case.    
 $V_X$ is a  homogeneous function of all moduli $\tau_i\equiv {\rm Re}(T_i)$ in general, but it is convenient to express $V_X$ in terms of 
Poincare dual coordinates $t_i$ related to 
$\tau_i$ by $\tau_i=\frac{\partial V_X(t_i)}{\partial t_i}$ : $V_X = V_X (\vec{t})$.  Note that, to compare to the $M$ theory case, $T_j \equiv -i \Phi_j$.
The degree of homogeneity of $V_X$ and the quantity $\gamma$ are related to the dimensionality 
of the internal manifold, $\gamma=2$ for Type IIB while $\gamma=3$ for $M$ theory. We will study the Type IIB case for concreteness but we will generalize the result at the end to include both cases.

In the above, the coordinates on the moduli space are the ``standard" supergravity ones $\tau_i$, but another equivalent choice is  ``polar coordinates", i.e. which divides the moduli 
space into a ``radial" piece and an ``angular" piece. The radial piece is parametrized by the overall volume $V_X$, while the other $N-1$ angular moduli are \emph{volume preserving}, 
denoted by $a_i \equiv \frac{\tau_i\,t_i}{V_X}$ . In other words, $a_i$ are homogeneous functions of $t_i$ of degree zero and changing their {\it vevs} does not change the overall
volume of the extra dimensions.
Secondly, not all $a_i$ are independent, since $\sum_i\,a_i=3/2$.

Now, it turns out to be very useful to consider the problem in the set of coordinates $\{a_i,\tau_D\}$, because they manifestly separate $\tau_D=Re(T_D)$ from the other moduli. These coordinates have the following interesting properties:
\begin{itemize}
\item The K\"{a}hler metric $K_{IJ};\;I,J=a_1,a_2,...,a_{N-1},\tau_D$ factorizes into a block diagonal form:
\ba \label{factor}K_{IJ} = \left( \begin{array}{cc}
K_{a_i a_j} & {\bf 0} \\
{\bf 0} & K_{\tau_D\tau_D}\\\end{array}\right)\ea 

\item From (\ref{WK}), it is easy to see that $\partial_{a_i}\,W=0,\;\forall i$, since $W$ only depends on $T_D$.
\end{itemize} 

We first study the properties of supersymmetric solutions of the system, i.e. $D_{T_D}\,W=0; D_{a_i}\,W=0;\,i=1,2...,N-1$. In particular, it can be shown that \cite{arXiv:1003.1982}: \ba\label{aisusy} D_{a_i}\,W|_{susy}&=&0;\;i=1,2...,N-1\nonumber\\ \implies \frac{\partial\,K}{\partial\,a_i}|_{susy} &=& 0\nonumber\\
\implies \; a_i^{susy} &=& n_i f(\{n_m,d_{jkl}\})\ea where $n_m$ and $d_{jkl}$ are constants determined by the topology of the compactification. 
The important thing to note is that the angular $a_i$ are completely determined by these constants!

For the $T_D$ modulus on the other hand, one gets:
\ba \label{tauDsusy} D_{T_D}\,W|_{susy} &=& 0\nonumber\\ \tau_D^{susy}&=& \frac{1}{b}\,\left[-\Omega_{-1}(\frac{-3|W_0|e^{-3/2}}{2A})-\frac{3}{2}\right]\nonumber\\ &=& \frac{1}{b}\,\left(\log(|\frac{2A}{3W_0}|)+...\right),\ea where $\Omega_{-1}$ is the non-principal branch of the Lambert-W function. The solution depends on the values of the ``microscopic" parameters - $\{A,b, W_0\}$.

Moreover, the overall volume of the manifold $V_X$ is completely determined in terms of the vev of $\tau_D$  and ``geometric" constants $\{n_m,d_{jkl}\}$ \cite{arXiv:1003.1982}:
\ba \langle V_X\rangle = \langle \tau_D \rangle^{3/2}\, \frac{1}{3^{3/2} (\sum_{jkl}\,d_{jkl}\,n_j\,n_k\,n_l)^{1/2}}\ea

We are ultimately interested in supersymmetry breaking minima with a positive vacuum energy. This generically requires additional non-moduli sources of supersymmetry breaking which provide a positive contribution to the superpotential. In principle, this can arise from various mechanisms in different situations -- a) $F$-term contributions from matter fields, b) $D$-term contributions from matter fields, or c) explicit sources of supersymmetry breaking, such as anti D-branes, etc. The important thing to note is that the qualitative result about moduli masses does not depend on the precise details as long as they satisfy the following criterion:
\begin{itemize}
\item The positive contribution to the potential only depends on the moduli through the overall volume $V_X$, to some (negative) power.
\end{itemize}

Most explicit models of phenomenologically relevant moduli stabilization satsify the above criteria, see the $M$ theory and Type IIB examples. The above feature gives rise to the following important consequences:
\begin{itemize}

\item   Since the vevs for $a_i$ in the susy extremum $a_i^{susy}$ are completely determined by the geometric constants, they are unchanged by the addition of the susy breaking contribution which depends on $V_X$!  

\item Only $\tau_D$ depends on $V_X$ (see (\ref{tauDsusy})), so its vev is affected by the presence of supersymmetry breaking. This then implies that the extremely complicated problem of minimizing the potential for a large number of moduli is replaced with the much simpler problem of minimizing the potential with respect to a single modulus $\tau_D$! 
\end{itemize}

Using these properties, it can be shown that the susy breaking vacuum obtained after including the positive contribution to the potential is ``close" (in field space) to the supersymmetric extremum above, i.e \emph{the vevs of the moduli in the true susy breaking vacuum above are close to that in the supersymmetric extremum}. In particular, \ba\tau_D^{true} &=&\tau_D^{susy} + {\cal O}\left(\frac{1}{b\tau_D}\right)^2\nonumber\\ a_i^{true}&=&a_i^{susy}\ea Note that the potential energy in the susy breaking vacuum can still be very different from that in the supersymmetric extremum.

Let us denote the additional positive contribution to the potential by $V_{lift}$, and for concreteness we use the example $V_{lift}=\frac{D}{V_X^n}$ with $n$ a positive number of ${\cal O}(1)$\footnote{For an explicit source of supersymmetry breaking, such as by an anti-Dbrane, $n=4/3$ or 2, depending on whether warping is present or not.}. This can be easily generalized to other cases. Using the fact that the true minimum is close in field space to the susy extremum, one can write: \ba \label{total-der}\frac{\partial^2\,V_{total}}{\partial\,a_i\partial a_j}|_{true} &\simeq& \frac{\partial^2\,V_{total}}{\partial\,a_i\partial a_j}|_{susy}\\ &\simeq& \frac{\partial^2\,V_{F}}{\partial\,a_i\partial a_j}|_{susy}+ \frac{\partial^2\,V_{lift}}{\partial\,a_i\partial a_j}|_{susy}\nonumber\ea Now using the fact that $V_{lift}$ depends on $V_X$ by a power law and that $\partial_{a_i}\,K|_{susy}=0$ (see (\ref{aisusy})), it is easy to show that: \ba \label{Vlift}\frac{\partial^2\,V_{lift}}{\partial\,a_i\partial a_j}|_{susy} = \frac{n}{\gamma}\,V_{lift}\,K_{a_i a_j}|_{susy}\ea But \ba\label{zerocc} V_{lift}|_{susy} \simeq 3\,m_{3/2}^2\ea due to the vanishingly small value of the cosmological constant.

The thing left to be computed is the first term in the second line in the RHS of (\ref{total-der}) - $\frac{\partial^2\,V_{total}}{\partial\,a_i\partial a_j}|_{susy}$. Since the K\"{a}hler metric factorizes as in (\ref{factor}), it is easy to see that the $F$-term potential $V_F$ is split into an ``$a_i$ piece" and a ``$\tau_D$ piece". Then, using the fact that $\partial_{a_i}\,W=0$ and $\partial_{a_i}\,K|_{susy}=0$, it is not hard to show that the Hessian matrix $\frac{\partial^2\,V_F}{\partial a_i\,\partial\,a_j}$, is: \ba \label{hessian_susy}\frac{\partial^2\,V_F}{\partial a_i\,\partial\,a_j}|_{susy} &=&   
-m_{3/2}^2\,K_{a_i\,a_j}|_{susy}\ea 

Using the results in (\ref{total-der}), (\ref{Vlift}), (\ref{zerocc}), and (\ref{hessian_susy}), the canonically normalized moduli mass-squared matrix at the true minimum, taking the effect of the non-trivial moduli K\"{a}hler metric into account, is then given by:
\ba \label{mass_true}{\cal M}_{ij}^2|_{true} &\simeq& \frac{1}{2}
(4K^{a_i\,a_k})\,(\frac{\partial^2\,V_{F}}{\partial a_k \partial a_j}|_{susy}+\frac{\partial^2\,V_{lift}}{\partial a_k \partial a_j}|_{susy})
\nonumber\\ &=& 2\,\left(3\frac{n}{\gamma}-1\right)\,m_{3/2}^2\,\delta_{ij}\ea which is true for both Type IIB and $M$ theory in general. $n$ and $\gamma$ are positive numbers of order one, so this proves the result ${\cal M}_{min}={\cal O}(1)\,m_{3/2}$. Of course, the values of $n$ and $\gamma$ which give rise to a (meta)stable vacuum are such that the ${\cal M}_{min}$ is positive. 

\subsection{Cosmological Consequences}

The above result that the lightest modulus mass is close to $m_{3/2}$ has a profound impact on pre-BBN cosmology. Current cosmological  data can only directly constrain the early Universe 
when it is colder than about a MeV, which is the onset of Big-Bang Nucleosynthesis (BBN). 
The most popular assumption for cosmological history before BBN is a ``thermal" history, i.e. in which the early Universe starts out with a radiation dominated phase due to reheating after 
inflation. In this case, the early Universe consists of a plasma of relativistic particles at a very high temperature.
The results obtained above, however,  question this assumption strongly under very general conditions.  
For a Hubble parameter during inflation which is bigger than about 30 TeV
\footnote{This is part of our set of assumptions, which seems quite natural. See section \ref{assumptions}.}, the moduli are displaced from their late-time minima during the early 
Universe, they start oscillating about their late-time minima when the Hubble parameter becomes comparable to their masses. Since they redshift like matter, they quickly dominate the 
energy density of the Universe giving rise to a matter-dominated phase. As mentioned earlier, the moduli interact gravitationally with all matter and hence have very long lifetimes. 
Requiring that the moduli decay reheats the Universe to temperatures above a few MeV thus puts a lower bound on their mass to be about 30 TeV. This gives rise to a ``non-thermal" 
cosmological history. The related gravitino problem is also solved in the following manner. The decay of the lightest modulus produces a lot of entropy, so the initial thermal abundance
 of the gravitinos is diluted away. Furthermore, the lightest modulus is lighter than 2\,$m_{3/2}$ in most examples\cite{arXiv:1003.1982,arXiv:0810.3285}, so that its branching ratio to 
gravitinos is also kinematically forbidden/suppressed \cite{Acharya:2008bk}, and a large abundance of gravitinos is not generated. 
  
A non-thermal history of the Universe before BBN has very important implications for many cosmological observables, 
and also for the origin and abundance of Dark Matter (DM). Before moving on to issues related to 
DM discussed in the next section in detail, we comment on possible cosmological observables which follow from the existence of a 
non-thermal cosmological history. One such observable could be the detection of gravitational waves produced during inflation, as pointed out in \cite{Durrer:2011bi}. Another observable is related to the growth of substructures in the early Universe. 
As shown in \cite{Erickcek:2011us}, the existence of a matter-dominated phase in the pre-BBN Universe leads to a significantly 
different pattern in the growth of structure. More studies are required to extract possible observable consequences.

\subsection{Range of $m_{3/2}$}

We saw that there is a lower bound on the gravitino mass of around 30 TeV from cosmological constraints. Is there also an upper bound on $m_{3/2}$ consistent with experimental constraints? The answer is yes for the following reason. 
We will see  that there exist compactifications in which gaugino masses are suppressed relative to $m_{3/2}$. In this case, the ratio of gaugino masses (at the unification scale $M_{GUT}$) to the gravitino mass is also determined in terms of the microscopic constants, and can be shown to be of the form: 
\ba M^a_{1/2} &\equiv& \frac{\sum_i^N\,F^i\partial_i\,f_{vis}}{2\,i\,{\rm Im}(f_{vis})}\\
  &\simeq& \frac{m_{3/2}}{{\cal O}(1)\,\log{\left(\frac{m_{pl}}{m_{3/2}}\right)}}\nonumber\ea 
where $F^i$ denotes the $F$-term the moduli labelled by $i$, $f_{vis}$ is the visible sector gauge kinetic function, and the ${\cal O}(1)$ number depends on the details of the $M$ theory \cite{hep-th/0701034,arXiv:1003.1982} or Type IIB examples \cite{hep-ph/0511162}. In the $M$ theory
case this can be seen, for example, by noting that the $F$-terms for the moduli are suppressed by ${\cal O}(1)\frac{\alpha_h}{4\pi} \sim \log(\frac{m_{pl}}{m_{3/2}})$ because susy
breaking is dominated by a hidden sector matter field and $\alpha_h$ measures the suppression factor. For more details, see section \ref{spectra}. This formula shows that if $m_{3/2}$ becomes larger than ${\cal O}$(100) TeV, then the gaugino masses become larger than a TeV or so, implying that the LSP again overcloses the Universe. Moreover, the axion relic abundance is proportional to a positive power of $m_{3/2}$ and naturally gives rise to 
an ${\cal O}(1)$ fraction of DM with minimal tuning only when $m_{3/2} \lesssim 100$ TeV \cite{Acharya:2010zx}, 
so this is another reason why one expects an upper bound on $m_{3/2}$ of around 100 TeV. 
We discuss these issues related to dark matter in more detail the following section. A point worth noting is that the upper limit on $m_{3/2}$ arises from
phenomenological input rather than theoretical constraints.

\section{Dark Matter}\label{dm}

A non-thermal cosmological history requires us to reassess our standard notions of DM vis-a-vis the nature of DM candidates and the parameter space of 
masses and interactions required to provide the entire DM content of our Universe. The two most attractive candidates for DM are the weakly interacting massive particle (WIMP), 
and the axion(s). We will find that the generic string/$M$ theory prediction is that both of these serve as excellent candidates and could each provide an 
${\cal O}(1)$ fraction of DM. Interestingly, the parameter space of masses and interactions required to provide the correct abundance is somewhat different
and, importantly,  much less fine-tuned compared to that with a standard thermal history. Let us discuss each of them in the following.

\subsection{WIMPs - Abundance, Signals for Direct and Indirect-detection}\label{wimp-dm}

The supersymmetric Standard Model with an exact or sufficiently conserved stabilizing symmetry (such as $R$-parity) naturally contains a WIMP DM candidate - 
the lightest superpartner charged under the particular symmetry, the LSP. 
The LSP in supersymmetric models with gravity mediation is typically a neutralino but other particles, such as staus or sneutrinos, are possible as well. 
However, as explained above using the result on the lightest modulus mass, one finds that the gravitino is generically heavier than around 30 TeV. 
This also generically sets the scale for squarks, sleptons and sneutrinos to be around 30 TeV. 
Gaugino masses, on the other hand, may or may not be suppressed relative to $m_{3/2}$ depending on the nature of moduli stabilization. 
Examples of both kinds of models exist in the literature. For compactifications with unsuppressed gaugino masses, see \cite{Cicoli:2008va}. 
The $\mu$ parameter, which determines the higgsino mass, 
can also be either of the same order or suppresed relative to $m_{3/2}$ depending upon the situation. 
See section \ref{spectra} for more discussion.

If gaugino masses are not suppressed then they, along with the squarks and sleptons, are too heavy to provide a viable WIMP DM candidate. In fact, this case can only be viable if the stabilizing symmetry of the LSP ($R$-parity) is violated so that the would-be LSP decays sufficiently rapidly. The case with suppressed gaugino masses is, therefore, significantly more interesting and can also be phenomenologically viable. This case can occur naturally in mechanisms of moduli stabilization in which the $F$-terms of moduli which determine the 
SM gauge couplings are suppressed relative to the dominant $F$ term, as in \cite{hep-th/0606262,hep-th/0701034,hep-ph/0511162,arXiv:1003.1982}. 
Then the lightest neutralino can be in the sub-TeV range, and can give rise to roughly the correct relic-abundance as follows. 
As explained above, the lightest modulus $X$ starts oscillating when $H \sim m_X$ and decays when $H = \Gamma_X$, the decay width of the modulus. 
The decay width of the modulus is given by: 
\ba\label{width}
\Gamma_X = \frac{D_X\,m_X^3}{m_{pl}^2}
\ea
where $D_X$ is a numerical coefficient which depends on the details of moduli stabilization and the compactification. 
With no prior knowledge, $D_X$ is generically assumed to be ${\cal O}(1)$ but in a given compactification, 
it can be larger, ranging from $\sim 10$ to $ \sim 10^4$ depending upon the micrsocopic details. 
A large $D_X$ is possible if the 
modulus {\it vev} measures the volume of a sub-manifold inside the internal manifold which is parametrically larger than unity in string or 11D length units. A large $D_X$ is equivalent to a modulus decay constant which is smaller than the Planck scale, 
$f_X = \frac{m_{pl}}{\sqrt{D_X}}$, i.e. it can be thought of as replacing the Planck scale in 
(\ref{width}) by a smaller scale such as the string scale or the compactification scale. It is a natural possibility, therefore, that the 
decay constants of the moduli are similar to those of the axions, which are also close to the compactification scale or the GUT scale, as will be seen in section \ref{axion-dm}\footnote{The precise values for the moduli and axion decay constants can differ by ${\cal O}(1)$ due to different mixing between the ``flavor" and mass eigenstates for the two cases.}.

The decay of the modulus reheats the Universe with a reheating temperature $T_R$ which has to be greater than a few MeV to satisfy BBN constraints.  
Since the canonically normalized $X$ is generally a linear combination of moduli whose values determine the gauge couplings, it has an ${\cal O}(1)$ coupling to gauginos. 
Thus, the branching ratio to the lightest neutralino is not small\footnote{Note that the modulus, being $R$-even, 
will decay to SM particles and superpartners, with all superpartners quickly cascade-decaying to the LSP. 
This will also add to the branching ratio to the LSP.}, 
giving rise to the number density of DM particles $\chi$ from $X$ decay as:
\be
n_{\chi}^X \sim \frac{\Gamma_X^2 \,m_{pl}^2}{m_{\chi}} \sim \frac{D_X^2\,m_X^6}{m_{pl}^2\,m_{\chi}}
\ee
This is to be compared with the critical density for annihilations of $\chi$-particles at the decay time ($H=\Gamma_X$):
\be
n_c \sim \frac{\Gamma_X}{\langle \sigma\,v\rangle} \sim \frac{D_X\,m_{X}^3}{m_{pl}^2\,\langle\sigma v\rangle}
\ee
For typical weak scale values of masses and cross-sections, $n_{\chi}^X$ is much larger than $n_c$, 
hence the DM particles annihilate after being produced until their number density becomes of order $n_c$. 
Thus, the final abundance of $\chi$ is given by: 
\be
n_{\chi} \sim n_c(T_{R}) \sim \frac{H(T_R)}{\langle \sigma v\rangle}\nonumber\\
\Omega_{\chi}\,h^2 \approx \Omega_{\chi}\,h^2_{(thermal)}\,\left(\frac{T_{F}}{T_R}\right)
\ee
where $T_F$ is the thermal freezeout temperature of the LSP. This gives rise to the following parametric dependence of ($\frac{\rho_{\chi}}{s} \equiv \frac{m_{\chi}\,n_{\chi}}{s}$) on the various quantities \cite{Acharya:2010af}: \ba \frac{\rho_{\chi}}{s} \simeq \frac{0.25\,\gamma_{\chi}}{D_X^{1/2}\,m_{3/2}^{1/2}\,m_{pl}^{1/2}\langle \sigma v\rangle},\ea which has to be normalized to the present value of the quantity,  $(\rho_{\chi}/s)_0 = 3.6\times 10^{-9}$ GeV to get the LSP relic abundance $\Omega_{\chi}\,h^2$. Here $\gamma_{\chi} \equiv \frac{m_{\chi}}{m_{3/2}}$ is the ratio of the low-scale LSP mass relative to $m_{3/2}$.

For a weak-scale LSP, $T_F$ is ${\cal O}(1-10)$ GeV while $T_R$ is ${\cal O}(1-10)$ MeV, 
giving rise to two-to-three orders of magnitude enhancement of the abundance over that of the thermal one for a given mass and cross-section. This implies that particles which naturally have a \emph{larger} cross-section
compared to that in the thermal case are good WIMP DM candidates. Hence, within the supersymmetric standard model, a wino-like LSP is a great candidate for DM in this context as it 
naturally provides the above enhancement in cross-section so as to roughly give rise to the correct abundance, see \cite{Acharya:2008bk, Acharya:2009zt}. 
A wino-like LSP can be naturally obtained in explicit models with moduli stabilization \cite{arXiv:0801.0478,arXiv:0810.3285}.  

If the above framework for Dark Matter is indeed realized in Nature, it can be constrained and tested by existing and future experiments.  Let us start with indirect detection. The data reported by PAMELA \cite{Adriani:2008zr} indicates an excess in the positron flux in the halo, which has been recently confirmed by FERMI \cite{FermiLAT:2011ab}. The annihilation of 140-200 GeV wino-like LSPs into a pair of $W$-bosons can naturally provide the excess observed in the positron flux with small boost (clump) factors \cite{Grajek:2008jb,Grajek:2008pg,Kane:2009if}. Contrary to what is naively thought, the reported PAMELA antiproton flux can also be consistent with wino annihilation \cite{Grajek:2008pg}.  The propagation model for positrons and antiprotons suffer from large uncertainties. It is, therefore, possible to vary the propagation model (consistent with all constraints) in a way so as to accommodate the antiproton signal and background with the reported antiproton flux. Data from the FERMI $e^+ + e^-$ and diffuse $\gamma$-ray flux from the whole sky  provide further constraints on the particle physics and astrophysics parameter space. For example, the FERMI $e^+ + e^-$ spectrum exhibits excesses from few tens of GeV to around a TeV. 
It is clear that a few hundred GeV wino LSP cannot account for both the PAMELA positron excess as well as the excess in the FERMI $e^+ + e^-$ data. 
Therefore, if the framework is correct this implies that there must be additional astrophysical sources of electrons and positrons \cite{Kane:2009if}. 
Such astrophysical sources could have many origins, as described in \cite{Zhang:2001ab, arXiv:0812.4457}. 
The diffuse $\gamma$-ray flux constraints in the relevant energy 
range arise from final-state radiation (FSR) $\gamma$-rays, which are satisfied by a wino-like LSP in the few 100 GeV range \cite{Papucci:2009gd}. 

Finally, stringent constraints arise from the FERMI search for gamma rays from DM-dominated dwarf spheroidal satellite galaxies of the Milky way. Since no DM signal has been found, upper bounds can be placed on the annihilation cross-section of WIMP DM candidates. In particular, a wino-like LSP with mass $\sim 200$ GeV annihilating into $W$-pairs is constrained to have $\langle \sigma v \rangle < 1\times 10^{-25}\,cm^3/s$ \cite{Garde:2011hd}. However, the annihilation cross-section of such LSPs from the theory turns out to be around an order of magnitude larger. Therefore, this implies that LSPs cannot account for the entire DM content of our Universe within this framework. The upper bound on  $\langle \sigma v \rangle$ is obtained \emph{assuming} that WIMPs constitute the enitre DM of the Universe. However, as we will see in section \ref{axion-dm}, axions naturally constitute an ${\cal O}(1)$ fraction of DM within our framework. This helps in relaxing the FERMI bound above since the signal scales as $\frac{J (\Delta\Omega)\,\langle \sigma\,v\rangle}{m_{\chi}^2}$ , where $J$ is the line-of-sight integral of the squared LSP relic density in the direction of observation over the solid angle $\Delta\Omega$: $J(\Delta\,\Omega) = \int_{\Delta\Omega}\,d\Omega\int_{l.o.s}\,dl\,\rho^2(l,\Omega)$. In particular, if  $\Omega_{\chi}\,h^2=\eta\,\Omega_{tot}\,h^2$, then the constraints are relaxed roughly\footnote{This is only a rough estimate, since the overall LSP abundance and the LSP density along the line of sight may be different in general.} by a factor of $\eta$. From a theoretical point of view, an LSP relic density which is an ${\cal O}(1)$ fraction smaller than the observed DM density can be obtained if the modulus decay constant $f_X$ is smaller than the Planck scale by around one-to-two orders of magnitude as explained below equation (\ref{width}). A comprehensive analysis of these issues with quantitative prospects for detection in various future experiments will appear in a forthcoming paper \cite{Kane:2012xx}.

To summarize, a wino-like LSP with a mass of around two hundred GeV is a viable DM candidate constituting an ${\cal O}(1)$ fraction of the DM relic abundance of the Universe. Future data from FERMI as well as AMS-02 \cite{Beischer:2009zz} will further clarify the situation. In order for the wino-like LSP DM candidate to remain viable with a non-negligible abundance, a mono-energetic $\gamma$-ray signal must be seen in the future by FERMI and/or AMS-02. Otherwise, the fraction of DM in the form of LSPs will be forced to be negligibly small.

Data from direct-detection and collider signals provide independent constraints. For the discussion of DM appearing in colliders, see section \ref{energy}. Here we briefly explain the prospects for direct-detection. It is well known that a pure wino LSP does not scatter against nuclei at tree-level because of the absence of Higgs and $Z$-exchange. Hence the interactions only arise at one-loop and are not strong enough to give rise to a detectable direct-detection signal in the near future. However, even with a small higgsino component ($\sim 10\%$), a wino-like LSP can give rise to a detectable direct-detection signal in next generation direct-detection experiments. String models generating a $\mu$ parameter consistent with such higgsino components exist, for the $M$ theory case see \cite{arXiv:1102.0556}.

\subsection{The ``Axiverse" - Abundance, Astrophysical signals}\label{axion-dm}

String compactifications to four dimensions generically give rise to a plethora of axions, as they reside in chiral supermultiplets along with the moduli fields.
Stabilizing the moduli with a sufficiently large mass so as to evade BBN constraints has interesting implications for axion physics. In  order for one of the axions to solve the strong CP-problem, i.e. to serve as the QCD axion, it should predominantly receive its mass from QCD instantons, and not additional stringy or supergravity effects. However, compactifications with moduli stabilized by only superpotential effects give axions masses comparable to $m_{3/2}$;\footnote{as in KKLT-type models \cite{Kachru:2003aw}.} hence none of the axions in these compactifications can solve the strong-CP problem. 

On the other hand, the $M$ theory and Type IIB moduli-stabilization mechanisms discussed above have moduli stabilized by a combination of K\"{a}hler potential and superpotential 
effects, as explained in detail in section \ref{mod-spectra}. In this case, only the moduli are stabilized at leading order while most of the axions are left unfixed. 
Integrating out the moduli and taking higher order effects into account, the axions are then  stabilized with masses exponentially suppressed relative to $m_{3/2}$ by higher order
non-perturbative effects. This gives rise to a spectrum of axions with masses distributed roughly evenly on a logarithmic scale \cite{Acharya:2010zx},  which was dubbed the ``Axiverse" in a more phenomenological approach \cite{Arvanitaki:2009fg}. In models with $M_{st} \gtrsim M_{GUT}$, one finds that tha axions can span a huge mass range from $m_a \sim H_0 \sim 10^{-33}$ eV to $m_a \sim 1$ eV. One of these light axions could naturally serve as the QCD axion if its mass is less than about $10^{-15}$ eV, hence solving the strong CP-problem \cite{Acharya:2010zx}.

One of the most important effects of these axions is their contribution to the total energy budget of the Universe. Axions start oscillating when $H \sim m_a$, and the energy in coherent oscillations could provide (at least part of) the dark matter of the Universe\footnote{Since there are many axions, they will start oscillating at different times. Also, the axions are so light that none of them have decayed yet.}. Within a thermal cosmological history, the WMAP bound on the relic abundance puts an upper bound 
on the axion decay constant $\hat{f}_a$ to be around $10^{11-12}$ GeV for an ${\cal O}(1)$ misalignment angle. On the other hand, with a non-thermal history, 
the computation of the axion relic abundance is different, and is schematically given by \cite{Fox:2004kb}:
\be
\Omega_{a_k}\,h^2 \simeq 10 \left(\frac{\hat{f}_{a_k}}{2\times 10^{16}\,{\rm GeV}}\right)^2\,\left(\frac{T_{R}}{10\,{\rm MeV}}\right)\,\langle\theta_{k}^2\rangle
\ee
for an axion $a_k$ which starts to oscillate in the moduli-dominated era. Here $\hat{f}_{a_k}$ is the axion decay constant, $T_R$ is the reheat temperature after the decay of the lightest modulus and $\langle\theta_{k}^2\rangle$ the average of the square of the initial {\it vev} of the axion when it starts oscillating (
for more details see \cite{Acharya:2010zx}). Therefore axions can
naturally give rise to the correct abundance for a much larger decay constant $\hat{f}_{a_k} \sim 10^{15}$ GeV and $T_R \gtrsim$ 5 MeV arising from the decay of a modulus heavier than around 25 TeV. With around 1 to 10$\%$ tuning of the misalignment angle (which may also arise from some hitherto unknown dynamical mechanism), $\hat{f}_a \sim M_{GUT} \approx 10^{16}$ GeV can also be accommodated. Within a GUT-motivated framework, this seems a much more natural possibility since 
the decay constant in string/$M$ theory solutions with unification tends to be around the GUT scale. One can view this both as a solution of the `cosmological axion decay constant problem' in string theory and also as a demonstration that axion physics is self-consitently much less fine-tuned with the non-thermal cosmological history generically predicted by string/$M$ theory than with a thermal one.

The Axiverse is subject to cosmological constraints and also has falsifiable predictions as discussed in \cite{Arvanitaki:2009fg, Acharya:2010zx}. For example, the observation of primordial gravitational waves would rule out the entire String-Axiverse under most conditions. Axions in the mass window $10^{-28} \lesssim m_{a} \lesssim 10^{-18} $ eV could give rise to step-like features in the matter power spectrum at small scales. On the other hand axions in the mass window $10^{-10} \lesssim m_a \lesssim 1$ eV can form bound states with black-holes, thereby significantly affecting their dynamics by graviton emission \cite{Arvanitaki:2009fg,Arvanitaki:2010sy}. It hardly needs to be emphasized that many interesting observables are possible and more studies are needed.

\subsection{Summary}\label{dm-summary}

To summarize, the arguments in the previous sections imply the following generic prediction (with suppressed gaugino masses) for Dark Matter:

\vspace{0.3cm}

Both WIMPs and axions are predicted together to form the DM content of our universe, each of them generically having a non-negligible fraction. This is possible if the modulus decay constant $f_X$ is around the string scale or GUT scale, hence smaller than the Planck scale by a few orders of magnitude. The precise fraction depends on the microscopic details of the compactification, and cannot be predicted with our current level of understanding. However, future experiments, especially for WIMPs, will be able to test this paradigm effectively and should either find firm evidence for it, or at least severely constrain it  if not exclude it conclusively. 

\vspace{0.3cm}

When gaugino masses are not suppressed, Dark Matter only exists in the form of axions.

\section{Particle Phenomenology}\label{particlepheno}

We now discuss important aspects of the broad phenomenology arising in the setup considered. We will elaborate on general features of the superpartner spectra discussed briefly in previous sections, followed by a discussion of aspects of electroweak symmetry breaking, and the  supersymmetric flavor and CP-problems.

\subsection{Superpartner Spectra}\label{spectra}

As discussed in section \ref{summary}, within the general setup considered, supersymmetry breaking in the hidden sector must be mediated to the visible sector by gravitational interactions, since otherwise there is a serious moduli problem. Thus, the gravitino mass sets the scale of all superpartners so one generically expects all the supersymmetry breaking mass parameters - the scalar masses, the trilinear parameters, and the gaugino masses to be ${\cal O}(m_{3/2})$. In particular, the expression for the soft scalar masses in ${\cal N}=1$ supergravity coupled to matter is given by \cite{Brignole:1997dp}:
\ba 
m_{\bar{\alpha}\beta}^2 = m_{3/2}^2 \tilde{K}_{\bar{\alpha}\beta} - \Gamma_{\bar{\alpha}\beta}
\ea where $\tilde{K}_{\bar{\alpha}\beta}$ is the K\"{a}hler metric for the matter fields $\alpha$ and $\beta$, and $\Gamma_{\bar{\alpha}\beta} \sim F^{\bar{i}}F^{j}\partial_{\bar{i}}\partial_{j}\tilde{K}_{\bar{\alpha}\beta}$ with $F^{i}$ as the $F$-term for the modulus or hidden matter field labelled by $i$.  Here $m_{pl}$ has been set to unity. The precise expression for $\Gamma_{\bar{\alpha}\beta}$ can be found in \cite{Brignole:1997dp}. In this basis the kinetic terms for the visible matter scalars are not canonical since $\tilde{K}_{\bar{\alpha}\beta}$ is non-trivial. To go to the canonically normalized basis, one does a unitary transformation ${\cal U}$ to make the K\"{a}hler metric diagonal, $({\cal U}^{\dag}\,\tilde{K}\,{\cal U})_{\bar{\alpha}\beta} = \hat{K}_{\alpha}\delta_{\bar{\alpha\beta}}$, and then does an appropriate rescaling for each field labelled by $\alpha$ to scale away the $\hat{K}_{\alpha}$. In the canonical basis, the mass-squared for the visible matter scalars is denoted by $\hat{m}_{\bar{\alpha}\beta}^2$ and is given by: \ba\label{can-scalar} \hat{m}^2_{\bar{\alpha}\beta}=m_{3/2}^2\delta_{\bar{\alpha}\beta}-\left(\frac{1}{\sqrt{\tilde{K}}}\,{\cal U}^{\dag}\Gamma\frac{1}{\sqrt{\tilde{K}}}\,{\cal U}\right)_{\bar{\alpha}\beta}\ea Since $\Gamma_{\bar{\alpha}\beta}$ depends on the derivatives of $\tilde{K}_{\bar{\alpha}\beta}$, the second term in (\ref{can-scalar}) is in general \emph{not} proportional to $\delta_{\bar{\alpha}\beta}$.  Both $\tilde{K}_{\bar{\alpha}\beta}$ and $\Gamma_{\bar{\alpha}\beta}$, however, depend on the values of the stabilized moduli, and are generically ${\cal O}(1)$ in string or 11D units. Hence, this gives rise to \ba \hat{m}_{\bar{\alpha}\beta}^2 = {\cal O}(1)\,m_{3/2}^2. \ea A similar analysis for the trilinears gives $\hat{A}_{\alpha\beta\gamma}={\cal O}(1)\,m_{3/2}$ in the canonically normalized basis.

Phenomenological models have been studied in which the above mass parameters are separated from the gravitino mass and/or from each other. For example, some models imagine that the supersymmetry breaking is ``sequestered" from the visible sector. Then tree-level contributions to scalar masses, gaugino masses and trilinears vanish and the dominant contributions arise from anomaly mediation  \cite{Randall:1998uk, Giudice:1998xp}. Similarly, models like split-supersymmetry \cite{hep-th/0405159} have studied scenarios in which the gaugino/higgsino masses are imagined to be protected by an $R$-symmetry which allows them to be vastly suppressed relative to the gravitino mass. 

What can be said about these scenarios from a string theory point of view? Let us start with the scalars and trilinears. As explained in section \ref{stabilize}, we discuss $M$ theory and Type IIB compactifications for concreteness. Within phenomenologically realistic $M$ theory compactifications which do not have any background closed string fluxes turned on, sequestering does not seem to be possible \cite{arXiv:0810.3285}. The situation in Type IIB string compactifications is more subtle. With partial moduli stabilization, it was argued that sequestering may be possible in the presence of strong warping \cite{Kachru:2007xp}, or due to the visible sector being localized in the extra dimensions \cite{Blumenhagen:2009gk}. However,after taking into account the stabilization of all moduli, there arise couplings between the moduli and the visible matter sector in the superpotential which do not allow for phenomenologically viable sequestering \footnote{``sort-of-sequestering", defined in \cite{Berg:2010ha}, may still be possible but that doesn't help.} \cite{Berg:2010ha}. Thus, we conclude that both scalar masses and trilinear parameters are generically of ${\cal O}(m_{3/2})$ in viable examples.

We have seen that scalar masses and trilinears are generically ${\cal O}(1)\,m_{3/2}$ in a general supergravity theory. In the $M$ theory and Type IIB moduli stabilization mechanisms considered, one can go much further since one can compute $\tilde{K}_{\bar{\alpha}\beta}$ and $\Gamma_{\bar{\alpha}\beta}$ in terms of the microscopic parameters. Moreover, these functions satisfy homogeneity properties at leading order in supergravity, and it can be shown that \cite{arXiv:0810.3285}: \ba \Gamma_{\bar{\alpha}\beta} \propto \tilde{K}_{\bar{\alpha}\beta} +\,{\rm higher\,order\, corrections} \ea
If these higher order corrections are small, as will be assumed in the following, then in the un-normalized basis one has:
\ba \label{scalars-M}
m_{\bar{\alpha}\beta}^2 &\simeq& m_{3/2}^2\,\left(1-\frac{7}{3}\frac{(m_{1/2}^{tree})^2}
{m_{3/2}^2}\right)\tilde{K}_{\bar{\alpha}\beta} \nonumber\\ &\simeq& m_{3/2}^2\,\tilde{K}_{\bar{\alpha}\beta},\ea which gives rise to the following in the canonically normalized basis: \ba \label{scalars-norm-M}
 \hat{m}^2_{\bar{\alpha}\beta} &\simeq& m_{3/2}^2\,\delta_{\bar{\alpha}\beta}\ea
where we have used the fact that $(m_{1/2}^{tree})^2 \ll m_{3/2}^2$ for these compactifications, and in the second line we have written the mass-squared matrix for the sfermion fields in the canonically normalized basis. Thus, scalar masses are very close to $m_{3/2}$ within the framework with stabilized moduli. A similar statement can be made for trilinears \cite{arXiv:0810.3285}.
  
What about gaugino masses? The situation in this case is different. Within gravity mediated supersymmetry breaking, which is preferred due to BBN constraints as explained in section \ref{assumptions}, $R$-symmetry is generically broken in the vacuum since (at least some) hidden superfields develop vevs for both their scalar and $F$-term components. In addition, for generic K\"{a}hler potentials arising in string compactifications, the anomaly mediation contribution is only one-loop suppressed relative to the gravitino mass \cite{Bagger:1999rd}. Hence, gaugino masses cannot be arbitrarily suppressed relative to the gravitino mass. Since $R$-symmetry is generically broken in the vacuum, the trilinear parameters are not suppressed and are of order $m_{3/2}$. The fact that both scalars and trilinears are of order $m_{3/2}$ will be crucial in mitigating the ``little hierarchy" which one would naively associate with such heavy scalars. This is discussed in the next subsection. Thus models which employ $R$-symmetry to suppress the gaugino masses (as in many split-supersymmetry examples) cannot arise within this framework. 

Although gaugino masses cannot be arbitrarily suppressed due to a \emph{symmetry}, they can still be somewhat suppressed relative to the gravitino mass by the \emph{dynamics} of moduli stabilization and supersymmetry breaking. This can be understood as follows. In many mechanisms of moduli stabilization, the geometric moduli which appear in the gauge kinetic function, $T^{vis}_i$, are stabilized ``close" to a supersymmetric point. The dominant supersymmetry breaking contributions which give rise to a dS vacuum are provided by other sources. Hence, the gaugino masses, which are proportional to the $F$-terms for $T^{vis}_i$, are suppressed relative to the gravitino mass in these situations.  This is true for the case of $M$ theory compactifications  \cite{hep-th/0606262,hep-th/0701034,arXiv:0810.3285} and was discussed in section \ref{stabilize-result}. In particular, using the results for moduli stabilization in section \ref{hierarchy-M} leads to a rather simple expression for the gaugino masses at tree level for phenomenologically viable cases: 
\ba
M_{1/2}^{tree}&=& \sum_{i=1}^{N}\,\frac{F^i\,\partial_i\,f_{vis}}{2\,i\,{\rm Im}(f_{vis})}\nonumber\\
&\simeq& -\frac{\alpha_h\,Q}{3\,\pi}\,m_{3/2}\left(1+{\cal O}(\alpha_h)\right) 
\ea where $F_i$ is the susy breaking $F$-term for moduli $i$ (the sum is over all $N$ moduli), and $f_{vis}$ is the visible sector gauge kinetic function which is an integer linear combination of all moduli, $f_{vis}=\sum_i^N\,{N^i\,s_i}$. $\alpha_h$ and $Q$ are defined in section \ref{hierarchy-M}, with the former related to the hidden sector gauge coupling $\alpha_h\equiv \frac{g_h^2}{4\pi}$, and $Q$ being an integer related to the rank of the hidden gauge group.  Note that the result is completely independent of the number of moduli $N$ as well as the integer coefficients $N^i$! As explained above, since $f_{vis}$ only depends on the moduli and not on the hidden field which is the dominant source of supersymmetry breaking, gaugino masses do not receive contributions from this dominant source and are hence suppressed relative to the gravitino mass. At one-loop, there are anomaly mediated contributions to the gaugino masses which turn out to be roughly of the same order \cite{hep-th/0606262,hep-th/0701034,arXiv:0810.3285}, hence they should be included as well.

Suppressed gaugino masses also arise in many classes of Type IIB compactifications \cite{Kachru:2003aw,hep-ph/0511162,arXiv:1003.1982}. However, within Type IIB compactifications, it could also happen that the $F$-terms for $T^{vis}_i$ are not suppressed, if, for example, they are stabilized by string-loop effects or perturbative effects in the K\"{a}hler potential \cite{Cicoli:2008va}. In these cases, the gaugino masses are expected to be of ${\cal O}(m_{3/2})$. However, as explained in section \ref{dm}, in this case the LSP abundance severely overcloses the Universe, so these are ruled out unless $R$-parity is sufficiently violated. Henceforth, we only discuss the case with suppressed gaugino masses. 

With suppressed gaugino masses $\lesssim$ TeV, the LSP can provide an ${\cal O}(1)$ fraction of DM with a non-thermal cosmological history if its annihilation cross-section $\langle \sigma\,v\rangle$ is a couple of orders of magnitude larger than the ``thermal"
annihilation cross-section ($\approx 3 \times 10^{-26}\,{\rm cm^3/s}$).  For further details,  refer back to section \ref{wimp-dm}. This can happen naturally if the LSP is wino-like or higgsino like with a mass of a few hundred GeV. Since gaugino masses are suppressed, anomaly mediation contributions turn out to be important. Then, it can be shown that natural choices of parameters can give rise to either a wino-like LSP or a bino-like LSP \cite{arXiv:0801.0478,arXiv:0810.3285}. However, a bino-like LSP has a much lower annihilation cross-section, hence its relic abundance will surely overclose the Universe in these theories. Hence, only wino-like LSPs seem to be possible if the LSP is gaugino-like. 

How much wino-like is the LSP? That depends  on the relative values of the wino mass parameter $M_2$, the bino mass parameter $M_1$ and the $\mu$ parameter which gives the higgsino mass. As demonstrated in \cite{arXiv:0810.3285}, for $M$ theory compactifications, it is possible to have $M_1$ and $M_2$ close to each other implying that the LSP can have a non-trivial  bino component. This would then decrease the annihilation cross-section of the LSP. A similar thing can happen in other string compactifications with suppressed gaugino masses as well. 

The magnitude of the $\mu$ parameter is more subtle and not yet settled in string theory. The phenomenologically viable value of $\mu$ in the canonically normalized basis of fields is around the TeV scale. It receives contributions both from supersymmetric terms in the superpotential and supersymmetry breaking terms arising from the K\"{a}hler potential. Hence, a natural option is to either forbid the $\mu$ term in the superpotential by a symmetry or have an approximate symmetry which suppresses the coefficient of the the holomorphic term $H_u\,H_d$ term in the superpotential by a large amount relative to the string/Planck scale (such as by exponential effects), and generate a viable $\mu$ term by K\"{a}hler potential effects - the Giudice-Masiero mechanism. However, the symmetry has to be such that large masses for color triplet fields \emph{are} allowed. Recall that  the spectrum is assumed  to arise from a GUT, hence the Higgs fields are part of a multiplet which also includes color triplet fields. These color triplet fields must get a large mass $\gtrsim M_{GUT}$ in order to not mediate proton decay at observably fast rates, which is the well-known ``doublet-triplet splitting problem".  Different string/$M$ theory soultions can contain different solutions to the doublet-triplet splitting owing to the different origin of matter and gauge degrees of freedom as well as differences in the underlying  structure of these compactifications. We are interested in those solutions which give rise to $\mu \leq m_{3/2}$. See \cite{hep-ph/9302227} for heterotic exampless, \cite{arXiv:0811.1583} for perturbative Type IIA and IIB cases, \cite{Heckman:2008qt} for $F$-theory and \cite{arXiv:1102.0556} for $M$ theory.  In $M$ theory compactifications one can further constrain $\mu$ by combining the requirements of moduli stabilization and the solution to the doublet-triplet splitting problem proposed by Witten \cite{Witten:2001bf}. This generically gives rise to $\mu \sim 0.1\,m_{3/2}$, but slightly smaller values may be possible as well. Thus, the LSP in this case will also have a higgsino component. If the $\mu$ parameter happens to be suppressed to an extent such that it becomes comparable to $M_2$, then it is possible for the LSP to have a significant higgsino component as well.

To summarize, the low-energy particle spectrum contains heavy scalars with masses $\gtrsim 30$ TeV. The gaugino masses may or may not be suppressed relative to the scalars depending on the microscopic details but viable low-energy theories only arise in cases where they are suppressed. In this case, it is possible to have wino-like LSPs which can naturally give rise to an ${\cal O}(1)$ fraction of DM. The LSP in general will have a bino and higgsino component as well, the precise amount depending on microscopic parameters i.e. more compactification specific details. 

\subsection{The ``Little" Hierarchy}\label{littlehierarchy}

It is well known that electroweak symmetry is broken by RG effects in a natural manner in the MSSM once one imposes soft supersymmetry breaking boundary conditions at around the unification scale. This is known as ``radiative electroweak symmetry breaking". This is because the RG equation for $m_{H_u}^2$ has a dependence on the top Yukawa coupling $y_t$ which 
is larger than all other Yukawa couplings. Hence, it is natural for $m_{H_u}^2$ to be driven to small or negative values, thereby destabilizing the  
point $H_u=H_d=0$ and giving rise to a Higgs vev. Thus, the higgs vev (or equivalently $m_Z$) becomes connected to the soft parameters and $\mu$. 

Although radiative EWSB is an extremely appealing feature of the MSSM, it turns out that obtaining the correct value of the $Z$ mass by choosing $O(1)$ values of soft parameters relative to a common scale $m_{soft}$ requires either a) $m_{soft} \sim m_Z$ or b) cancellation between soft parameters (essentially $m_{H_u}^2$ and $\mu^2$ when $\tan\beta$ is not small) of order $m_{soft}^2$, with $m_{soft}$  larger than $m_Z$. The former option turns out to be incompatible with direct constraints on superpartner masses as well as the Higgs mass bounds from LEP, leaving the latter as the only option. This is the infamous ``little hierarchy" problem in the MSSM.  Note that this is not just true for the MSSM, the bounds on masses of new physics particles from direct production as well as bounds from indirect electroweak precision data imply that the problem is generically present in all other approaches to electroweak symmetry breaking such as warped extra dimensional models, composite higgs models, little higgs models, etc. and to a lesser extent even in weakly coupled models with an extended matter sector such as the NMSSM.

Since the framework considered here assumes the MSSM matter and gauge spectrum, the fact that the scalar superpartners are heavier than around 30 TeV would naively seem to suggest a much more severe fine-tuning compared to MSSM models with scalar masses $\lesssim$ TeV. However, this turns out to be incorrect in models where $\mu$ is also suppressed, as we explain below. The basic reason is that in gravity mediation with no sequestering of the visible sector fields relative to the hidden sector, which seems to arise naturally within string theory solutions providing a solution to the moduli problem, \emph{both} scalar masses ($M_0$) and trilinears ($A_0$) are close to each other, of ${\cal O}(m_{3/2})$. Since $M_0$ and $A_0$ appear in the RG equation for the Higgs mass-squared parameter $m_{H_u}^2$ with opposite signs, this gives rise to a near cancellation between the two terms, giving rise to a  $m_{H_u}^2$ which is naturally suppressed relative to $m_{3/2}^2$ \cite{arXiv:1105.3765}. More concretely, $m_{H_u}^2$ at any given scale $Q$, is given as a function of $t \equiv \log(Q/Q_0)$, with $Q_0$ the unification scale, by: 
\be \label{mhu2}
m_{H_u}^2(t) \simeq f_{M_0}(t)\,M_0^2-f_{A_0}(t)\,A_0^2 + R(t)
\ee
The quantities $f_{M_0}$ and $f_{A_0}$ are determined by SM Yukawa couplings and gauge couplings at leading order. $R$, on the other hand, is determined primarily by the gluino mass parameter $M_3$ and hence gives a negligible contribution if $M_0, A_0 \gg M_3$, as is the case here. Then one finds that for $M_0 \simeq A_0 \simeq m_{3/2}$, $f_{M_0}$ and $f_{A_0}$ at the electroweak scale are naturally of order 0.1 and also nearly cancel each other \cite{arXiv:1105.3765}, implying that:
\be
m_{H_u}^2 (Q_{EWSB}) \sim 10^{-2}\,m_{3/2}^2 \sim {\rm TeV}^2
\ee
Thus, in compactifications where $\mu$ is ``small', i.e. $\mu^2 \sim 10^{-2}\,m_{3/2}^2$, the naive fine-tuning is significantly reduced. For more details, please refer to \cite{arXiv:1105.3765}. Note that this mechanism, dubbed the ``Intersection-Point" in \cite{arXiv:1105.3765},  is quite different from the ``Focus-point" region in the constrained MSSM \cite{hep-ph/9908309}, where $A_0$ at the unification scale is much smaller than the large soft mass parameters. When $\mu$ is ``large", i.e. of the same order as $m_{3/2}$, the fine-tuning is quite severe as expected.

Even for small $\mu$, since $m_{H_u}^2 \sim {\rm TeV}^2$ rather than $m_Z^2 \sim 100\,{\rm GeV}^2$, it appears that some degree of fine-tuning, at least from an electroweak scale point of view, still remains. From a top-down point of view, however, two possibilities exist. While it may be possible that a fine-tuning is intrinsically present, it is also possible that the fine-tuning is ``apparent" and is just a manifestation of our less-than-perfect understanding of the underlying theory at the high scale. For example, for small $\mu$ which is around two orders of magnitude suppressed relative to $m_{3/2}$, \emph{and} for $A_0$ very close to 1.2 $M_0$, $m_{H_u}^2$ can be as low as ${\rm few}\times 10^{-4}\,m_{3/2}^2$ (see Figure 3 in \cite{arXiv:1105.3765}) ! So, if there is an underlying (unknown) reason for such a value of $A_0/M_0$, then the fine-tuning does not exist at all. It is important to be aware of such a possibility. Finally, it is interesting to note that in a different context it has been argued that no physics would change if the higgs vev, which is equivalent to $m_Z$, were several times larger than the experimental value \cite{Donoghue:2009me}.

\subsection{The Higgs Mass}

 The prediction of the Higgs mass is one of the most important predictions of any framework beyond the Standard Model, hence it is important to understand what our framework has to say about the Higgs mass, which is the topic of this subsection. We note that the prediction was first presented at the International String Phenomenology Conference at the University of Wisconsin, Madison in August 2011. A more refined prediction appeared on the arXiv in \cite{Kane:2011kj}. It is remarkable that the recent hints of the Higgs signal by ATLAS and CMS at around 125 GeV are naturally consistent with the prediction. We now explain the relevant details of the Higgs mass computation.

\begin{figure}[h!]
\includegraphics[width=3.45in,height=2.9in]{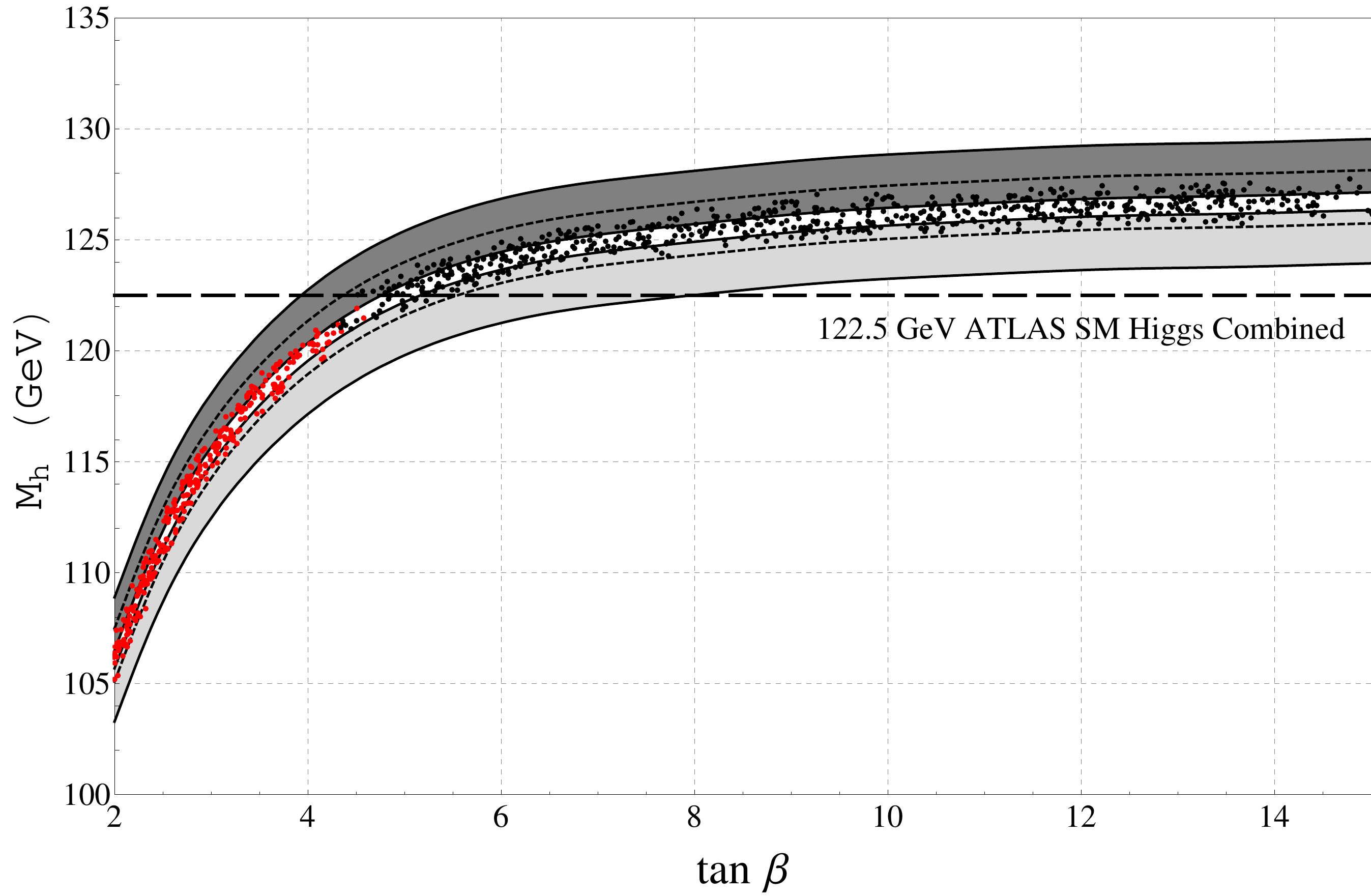}
\caption{\small{The prediction for the Higgs mass at two-loops for realistic string/$M$ theory vacua defined in the text, as a function of $\tan \beta$ for three different values of the gravitino mass $m_{3/2}$, and varying the theoretical and experimental inputs as described below. For precise numbers and more details, see \cite{Kane:2011kj}. The central band within the dashed curves for which scatter points are plotted corresponds to $m_{3/2}=50$ TeV. This band includes the total uncertainty in the Higgs mass arising from the variation of three theoretical inputs at the unification scale, and from those in the top mass $m_t$ and the $SU(3)$ gauge coupling $\alpha_s$ within the allowed uncertainties. The innermost (white) band bounded by solid curves includes the uncertainty in the Higgs mass for $m_{3/2}=50$ TeV only from theoretical inputs. The upper (dark gray) band bounded by solid curves corresponds to the total uncertainty in the Higgs mass for $m_{3/2}=100$ TeV while the lower (light gray) band bounded by solid curves corresponds to that for $m_{3/2}=25$ TeV.  For $m_{3/2}=50$ TeV, the red scatter points (with $\tan\beta$ less than about 4.5) and blue scatter points (with $\tan\beta$ greater than about 4.5) correspond to ``Large" $\mu$ and ``Small" $\mu$ respectively, as described in the text.}}
\end{figure}

First, since supersymmetric models require two Higgs doublets for anomaly cancellation, by the ``Higgs mass" it is meant the mass of the lightest CP-even neutral scalar in the Higgs
sector. A remarkable fact about the Higgs mass even in
general supersymmetric theories is that an upper limit on $M_h$ of order $2\,M_{Z}$
exists just from the requirement of validity of perturbation
theory up to the GUT scale \cite{hep-ph/9210242}. 
This is due to the
fact that the Higgs mass at tree-level only depends on SM gauge couplings
(which have been measured), and possibly other Yukawa or gauge couplings
(which are bounded from above by perturbativity). However, in addition to
the gauge and matter spectrum, the precise value of the Higgs mass depends
crucially on radiative effects, which in turn depend on all the soft
parameters including the $\mu $ and $B\mu $ parameters.

Since the generic string theory scalar superpartner masses are much larger than the electroweak scale and viable models have gaugino masses which are $\lesssim$ TeV, it is useful to integrate out the scalars below their characteristic mass scale ($\gtrsim 30$ TeV), and study the effective theory consisting of the Standard Model particles, charginos and neutralinos. As discussed earlier, higgsinos may or may not be suppressed depending on the value of $\mu$, we consider both cases when computing the Higgs mass. Note that all Higgs scalars except the lightest CP-even Higgs $h$ are quite heavy, with masses close to $m_{3/2}$. Thus, we are in the so-called ``decoupling limit" of the two-Higgs doublet model. In this case, the Higgs mixing angle\footnote{for its definition, please refer to section 8.1 in \cite{Martin:1997ns}} $\alpha$ is related to $\beta$ as $\alpha = \beta - \frac{\pi}{2}$, and the lightest CP-even Higgs behaves very close to the Higgs in the Standard Model.

The lightest CP-even Higgs mass, $M_h$, is given by: $M_h = \sqrt{2\,\lambda}\,v$, where $\lambda$ is the Higgs quartic coupling and $v=174$ GeV is the Higgs vev. In the MSSM, $\lambda = \frac{g^2+g'^2}{8}\,\cos^2(2\beta)$ at tree level\footnote{$\beta$ is defined as $\beta \equiv \tan^{-1}(\frac{v_u}{v_d})$ where $v_u$ and $v_d$ are the vevs of the two Higgses in the supersymmetric two-Higgs doublet model.}, which is small. Once the squarks, sleptons and heavy Higgs scalars are integrated out around their mass-scale, this gives rise to a threshold correction to the quartic coupling $\delta\lambda$. In addition, there are corrections arising from loops of supersymmetric fermions at around the electroweak scale, which we denote by $\delta \tilde{\lambda}$. The lightest Higgs mass $M_h$ is thus given by:
\be
M_h = \sqrt{2}\,v\,\sqrt{\lambda+\delta\lambda  +\delta\tilde{\lambda}}
\ee
The corrections $\delta{\lambda}$ and $\delta\tilde{\lambda}$ also depend on $\tan\,\beta\equiv \frac{v_u}{v_d}$ in general. For details on the Higgs mass computation, see \cite{Kane:2011kj} and references therein.

It is important to understand the dependence on $\tan\beta$. Theoretically, $\tan\beta$ is not a free parameter, it is determined by the soft terms and $\mu$. However, since $\mu$ and $B\mu$ may or may not be suppressed relative to $m_{3/2}$ depending upon the microscopic details, it is possible to use $\tan\beta$ as an input and then determine $\mu$ and $B\mu$ by the electroweak symmetry breaking condition. However, only those values of $\tan\beta$ are chosen which yield $\mu$ and $B\mu$ consistent with theoretical expectations. A consequence of this is that the values of $\mu$ and $\tan\beta$ are correlated : $\tan \beta \gtrsim 4.5$ is possible only when $\mu$ is suppressed relative to $m_{3/2}$, while $\tan\beta \lesssim 4.5$ is possible only for unsuppressed $\mu$. Using  these ingredients, the Higgs mass can be computed as a function of $m_{3/2}$, see Figure 1. We see from the Figure that the Higgs mass naturally lies between 105 GeV and 129 GeV depending on $\tan\beta$. For small $\mu$, the prediction is more refined: $122 \lesssim M_h \lesssim 129$ GeV \cite{Kane:2011kj}.

The Higgs mass prediction holds for all compactifications with an MSSM matter and gauge spectrum below the compactification scale and with scalars heavier than around 25 TeV and gaugino masses $\lesssim$ TeV. In addition to the dependence on $\tan\beta$, there is a mild dependence on the overall scale $m_{3/2}$ which we allow to vary from $\sim 25$ TeV to $\sim$ 100 TeV. For a given $m_{3/2}$ and $\tan\beta$, there is a small spread in the Higgs mass prediction arising from variation of theoretical inputs like the trilinears and the gluino mass parameter within reasonable ranges at the GUT scale, as well as from experimental uncertainties in the top mass and the strong gauge coupling. A precise experimental value for the Higgs mass will constrain $m_{3/2}, \mu$ and $\tan\beta$ significantly.

Note that other authors have earlier proposed that interpreting data 
in the context of supersymmetry (as well as a variety of different theoretical assumptions)
was suggestive of scalars heavier than
might have been naively expected \cite{Wilczek}, \cite{hep-th/0405159}.
The string/$M$ theory
derivation leads to a definite scale  for the scalar masses (tens of TeV at the
unification scale) which gives a fairly sharp prediction for the Higgs mass.

To summarize, the framework predicts that ATLAS and CMS should report conclusive evidence for the Higgs this year. It also makes precise predictions about Higgs properties. Since we are in the decoupling limit of the Higgs sector in the MSSM, the Higgs behaves very similarly to the SM Higgs. In particular, the Higgs production cross-section in the gluon fusion channel is virtually indistinguishable from that in the SM due to the stops being much heavier than the tops. Furthermore, including the effects of superpartrners,  the branching ratios to  $b\,\bar{b}$ (which dominates the total width) and other modes such as $\gamma\,\gamma$, $Z\,\gamma$ do not deviate from the SM by more than a few percent. Hence, it is predicted that the Higgs will be virtually indistinguishable from the SM Higgs, at least in the near future.

\subsection{Flavor and CP}\label{susy-flavorcp}

The origin and pattern of the quark and lepton Yukawa couplings in the Standard Model still remains a mystery, although from a top-down point of view progress has been made in understanding at least some of the issues involved \cite{Heckman:2008qa, Leontaris:2010zd}. In this work, we will not discuss the origin of fermion flavor, their masses and mixings. These issues are addressed by string theory but not yet resolved. Here we will be mainly interested in flavor and CP issues related to the beyond-the-SM (superpartner) sector. While studying these issues, inputs about fermion masses and mixings (excluding that of neutrinos) will be needed at various places; we will assume that the Yukawa couplings are hierarchical 
since this is generically true in string/$M$ theory; for instance, in $M$ theory compactified on a $G_2$ manifold chiral fermions and Higgs fields are localized at different points in the extra dimensions and
the Yukawa couplings are generated by membrane instantons and are exponentially suppressed by the volumes of 3-cycles which give the instanton action. Small variations in
these volumes then lead to large hierarchies \cite{Acharya:2001gy,Acharya:2003gb}. See \cite{arXiv:0811.1583} for analagous examples in string theory.

It is well known that \emph{without} any underlying structure, gravity mediation models generically lead to too large predictions for flavor and CP-violating observables. However, within the context of an underlying supersymmetry breaking mechanism arising in string theory, additional underlying structures, which help shed more light on these issues, are often present. One possibility is that the underlying string compactification preserves flavor symmetries which could effectively suppress flavor and CP-violation \cite{Ross:2002mr}. Another possibility arises from the structure of the underlying hidden sector and moduli dynamics associated with supersymmetry breaking and its mediation to the visible sector. We will focus on the latter as this is more directly connected to the physics of moduli stabilization. But it is important to be aware that the presence of flavor symmetries will further help in constructing viable models consistent with all phenomenological constraints. Also, for concreteness, we will focus on the $M$ theory case as these issues have been studied in detail in this context. However, we expect the qualitative results to hold for other classes of compactifications as well, such as the class of Type IIB compactifications described in section \ref{other}.

The relevant soft parameters to consider are the scalar mass-squared matrices $m_{\bar{\alpha}\beta}^2$ and the trilinear parameters $\tilde{A}^{\gamma}_{\alpha\beta} \equiv A_{\gamma\alpha\beta}\,Y^{\gamma}_{\alpha\beta}$, where $\alpha,\beta$ stand for the flavor indices and $\gamma$ stands for the up-type or down-type sector. 
$Y^{\gamma}_{\alpha\beta}$ are the normalized Yukawa couplings of the fermions. In a general ${\cal N}=1$ supergravity theory, the expressions for the scalar mass-squared matrix (with vanishing cosmological constant) is given by \cite{Brignole:1997dp}:
\be\label{softmass}
m_{\bar{\alpha}\beta}^2 = m_{3/2}^2 \tilde{K}_{\bar{\alpha}\beta} - \Gamma_{\bar{\alpha}\beta}
\ee
where $\tilde{K}_{\bar{\alpha}\beta}$ is the K\"{a}hler metric for the matter fields $\alpha$ and $\beta$. $\Gamma_{\bar{\alpha}\beta} \sim F^{\bar{i}}F^{j}\partial_{\bar{i}}\partial_{j}\tilde{K}_{\bar{\alpha}\beta}$ with $F^{i}$ as the $F$-term for the modulus or hidden matter field labelled by $i$. The precise expression can be found in \cite{Brignole:1997dp}. Without any underlying structure, the second term in (\ref{softmass}) gives rise to ${\cal O}(1)$ flavor violation even when one goes to the canonically normalized basis where $\tilde{K}_{\bar{\alpha}\beta}$ in the first term is rotated away. This is because $\Gamma_{\bar{\alpha}\beta}$ involves derivatives of $\tilde{K}_{\bar{\alpha}\beta}$ which are not proportional to $\tilde{K}_{\bar{\alpha}\beta}$ in general. However, within realistic moduli stabilization mechanisms as described in section \ref{stabilize}, $\tilde{K}_{\bar{\alpha}\beta}$ satisfies homogeneity properties at leading order that are broken by higher order derivative corrections. It is very hard to compute these higher order corrections from the microscopic theory. We assume that these higher order corrections are either small and/or have approximately the same flavor structure as that for the leading order\cite{arXiv:0810.3285}, see also the discussion in section \ref{spectra} above eqn. (\ref{scalars-M}). 

Using these, it can be shown that $\Gamma_{\bar{\alpha}\beta}$ is proportional to $ \tilde{K}_{\bar{\alpha}\beta}$ to a good approximation. This gives rise to approximately flavor-diagonal and universal soft mass-squared matrices, as shown in \cite{arXiv:0810.3285}.  Note, however, that this is only true at the scale where the boundary conditions for the RG evolution are imposed, in this case the unification scale. RG effects and rotation to the super-CKM basis in general introduce a small amount of  flavor violation. This can be parametrized by the quantities :
\be
(\delta_{XY})_{\alpha\beta} = \frac{(\hat{m}_{XY}^2)_{\alpha\beta}}{\sqrt{(\hat{m}_{XY}^2)_{\alpha\alpha}(\hat{m}_{XY}^2)_{\beta\beta}}}
\ee
Here $X, Y \in \{L,R\}$ and a hat denotes a matrix in the super-CKM basis.

The moduli stabilization mechanism discussed in sections \ref{mtheory} and \ref{other} also has an important consequence for the supersymmetric (weak) CP problem. An important feature of the moduli stabilization mechanism is that it gives rise to a \emph{real} superpotential in the vacuum  at leading order, i.e. it does not contain any CP violating phases. This has important consequences for CP violation in the flavor diagonal and off-diagonal sector as we will see.  The reason is as follows. As explained in section \ref{mod-spectra}, at  leading order the potential stabilizes all the moduli but only stabilizes a few axions. 
For example, for the case considered in detail in section \ref{mod-spectra}, there are two terms in the superpotential at leading order. Then, it can be shown that one axionic combination $t$ is stabilized such that $\cos\,t=-1$, implying that the terms in the superpotential align with the same phase (apart from a sign) \cite{Acharya:2010zx,Kane:2009kv}. Since the overall phase of the superpotential can be rotated away and is not observable, this means that the superpotential in the vacuum is real at leading order. As mentioned in section \ref{axion-dm}, all remaining axions are stabilized by effects by other non-perturbative terms in the superpotential which are exponentially suppressed relative to the leading terms\footnote{This can happen quite naturally since the arguments of these exponential terms are essentially given by the volume of sub-manifolds. So, if these volumes are just ${\cal O}(1)$ larger than those in the leading exponential, these terms will be highly suppressed.}, making them exponentially lighter than the gravitino mass and hence solve the strong CP-problem. The same also implies that once these remaining axions are stabilized, there may be terms in the superpotential with different phases, however since these terms are exponentially suppressed relative to the leading terms, they can be neglected to an excellent approximation. {\it It is worth emphasising that the solutions to both the weak and strong CP problems have a common origin}.

Using the above, one can show that the soft supersymmetry breaking parameters in the Lagrangian are real at the unification scale to an excellent approximation \cite{Kane:2009kv}. This implies that in particular the gaugino masses and reduced trilinears $A^{\gamma}_{\alpha\beta} \equiv \tilde{A}^{\gamma}_{\alpha\beta}/Y^{\gamma}_{\alpha\beta}$ are real as well.  Using the homogeneity properties of $\tilde{K}_{\bar{\alpha}\beta}$, it is possible to show that $\tilde{A}^{\gamma}_{\alpha\beta}$ is roughly proportional to Yukawa couplings at the unification scale. Again, RG effects and rotation to the super-CKM basis introduce CP phases in the trilinear parameters in both the flavor-diagonal and off-diagonal sector. The most stringent constraints arise from observables like $\epsilon_K$, ${\rm Re}(\epsilon'/\epsilon)$ and electric dipole moments (EDMs) \cite{Kadota:2011cr}. CP-violation in the flavor off-diagonal sector affects observables like $\epsilon_K$ mainly through chirality-conserving interactions, while that in the flavor diagonal sector affects EDMs through chirality-flipping interactions. The real part of $\epsilon'/\epsilon$ gets dominant contributions from chirality-flipping flavor-violating effects such as $(\delta_{LR})_{12}$ and $(\delta_{RL})_{12}$. Utilizing the properties mentioned above, the contributions to all the above flavor and CP-violating observables were computed in detail in \cite{Kane:2009kv,Kadota:2011cr}, and it was found that all such constraints are satisfied with hierachical Yukawa couplings, with scalar masses and trilinears $\gtrsim 30$ TeV, and with gaugino masses $\lesssim$ TeV. Predictions were also made for various EDM measurements in \cite{Kane:2009kv}. In section \ref{precision}, we discuss possible experiments at the precision frontier which could test and constrain the framework. 

Note that there is a qualitative difference between the electron and hadronic EDMs. The former is virtually vanishing in the SM, but the latter does receive a contribution from the $\theta$ angle in QCD, which in fact is the origin of the strong CP-problem. Therefore, once EDMs are observed for the electron, neutron, mercury, etc., it will be important to separate the suspersymmetric (BSM) contribution from the contributions proportional to $\theta$. Note that in the solution to the strong CP problem described in section \ref{axion-dm}, the value of $\theta$ is in principle determined by microscopic constants which also affect astrophysical observables. So, this gives rise to an extremely interesting (albeit indirect) connection between astrophysics and precision observables, which should be explored further.

\section{High Energy Physics Signals}\label{hep-signals}

It is natural to ask how the framework we have studied manifests itself at high energy physics experiments. These can be broadly divided into two categories - the energy frontier and the precision frontier. We discuss both of these below.

\subsection{Energy Frontier}\label{energy}

The LHC has achieved significant milestones in its performance and has amassed a wealth of high-quality data.  It has already ruled out a significant region of parameter space of many beyond-the-Standard-Model frameworks. What can be said about the framework studied here in terms of LHC signals? Since scalars are heavier than about $30$ TeV, they cannot be directly produced at the LHC. However, 
the framework predicts gaugino masses $\lesssim$ TeV, so they should be accessible at the LHC. Therefore, the most promising channel is pair production of gluinos followed by their decay to a realtively high mutliplicity of third generation fermions such as top and/or bottom quarks. The gluinos have a large production cross-section because they carry color and are fermions. However, their cross-section is suppressed relative to the case with comparable squark masses. The gluinos decay via virtual squarks into $q\bar{q}\chi_1^0$ or $q\bar{q}\chi_1^{\pm}$ since the squarks are heavier than the gluinos. Since the rate scales as $m_{\tilde{q}}^{-4}$, the lightest squarks dominate the process.  If all the scalar masses are roughly equal at the unification scale (close to $m_{3/2}$), then RG effects drive the third generation squarks to be lighter than the first two. Thus, the gluino decay channels $\tilde{g} \rightarrow t\bar{t}\chi_1^0,\, t\bar{b}\chi_1^{\pm}, \, b\bar{b}\chi_1^0$ dominate over a large region of parameter space\footnote{Gluino decays to $\chi_1^{\pm}q\bar{q}'$, and $\chi_1^0q\bar{q}$ are also significant.}. These lead to $b$-rich and lepton-rich final states with excellent prospects for discovery. Detailed studies of these kind of models have been carried out for the 14 TeV LHC in \cite{Acharya:2009gb}, and for the 7 TeV LHC in \cite{Kane:2011zd}. LHC studies of phenomenological models with a similar spectrum have also been performed in \cite{Baer:2011sr}. In particular, for the 7 TeV LHC, it has been shown that the 1 lepton channel with at least four $b$-tagged jets is particularly sensitive to this class of models even with moderate amounts of data. 
In some cases, the same-sign (SS) dilepton channel can also be a competitive model for discovery since it encounters fewer backgrounds from SM processes. 

What about signals of the chargino and neutralino sector? As explained in section \ref{wimp-dm}, the LSP in the framework is wino-like with a small bino component. The higgsino component depends on the value of $\mu$ and could be either small or significant. The lightest chargino $\tilde{\chi}_1^{\pm}$ and the lightest neutralino $\tilde{\chi}_1^0$ are quasi-degenerate if the LSP is mostly wino, with $m_{\tilde{\chi}^{\pm}_1}-m_{\tilde{\chi}^{0}_1} \lesssim 200$ MeV. In this case, the charginos decay to the LSP emitting very soft pions or leptons. 
Thus gluino decays to charginos can lead to the  charginos traveling through two or three layers of the tracker and then decaying, giving rise to disappearing high $p_T$ charged tracks. Observing this signal is challenging and requires a dedicated analysis, but should be possible \cite{Kane:2012aa}. Electroweak production of charginos and neutralinos has a significant cross-section \cite{Acharya:2009gb} and should also be observable eventually, and can help provide experimental information about the nature of the LSP. The tree-level production for  $\tilde{\chi}_1^{\pm} +  \tilde{\chi}_2^{0}$ vanishes for a pure bino LSP, so the cross-section is sensitive to the bino component of the LSP. Similarly, the rate for production of   $\tilde{\chi}_1^{\pm} +  \tilde{\chi}_1^{0}$ is about two times larger for a wino LSP than for a higgsino LSP and can thus help determine the LSP type.

To summarize, this set of ideas gives rise to many falsifiable predictions at the LHC which are being probed currently. For instance, the analysis presented by ATLAS \cite{ATLASb-jet} puts a lower limit on the gluino mass of roughly 700 GeV for a neutralino mass of 150 GeV. Most importantly, gauginos must be eventually observed at the LHC with enhanced branching fractions to the third generation else the case with suppressed gaugino masses will be ruled out.

\subsection{Precision Frontier}\label{precision}

Precision mesurements are sensitive to new particles running inside loops, and hence can indirectly probe BSM physics. Within our framework, the fact that the scalar masses and trilinears are $\gtrsim 30$ TeV while the gauginos are $\lesssim$ TeV helps keep the flavor and/or CP violating effects under control, as explained in section \ref{susy-flavorcp}. However, some  measurements can be sensitive to new physics within the framework in the near future. For example, an improvement in the constraints from $b\rightarrow s\,\gamma$ by about an order of magnitude will start probing the framework \cite{Kadota:2011cr}. Similarly, EDM predictions from the framework naturally turn out to be one-to-two (for the mercury EDM), two-to-three (for the neutron EDM), and about four (for the electron EDM) orders of magnitude smaller than the current limits \cite{Kane:2009kv}. So, an improvement in these limits in the future will be able to test the framework. In addition, even though the decay width of  $B_s \rightarrow \mu\,\mu$ is proportional to $\tan\beta^6$, we find that the prediction is still virtually indistinguishable from the SM (Our current understanding of the theory suggests $\tan \beta \lesssim 20 $ \cite{arXiv:1102.0556}).

\section{The Matter-Antimatter Asymmetry}\label{matter-asymm}

Finally, we discuss the origin of the matter-antimatter asymmetry of the Universe within the framework and its connection to the LSP abundance. The presence of light moduli imply a period of moduli domination shortly after the end of inflation. This era lasts until the lightest modulus decays providing a reheating temperature high enough for successful nucleosynthesis. However, the decay also produces a large amount of entropy greatly diluting any pre-existing abundances in the Universe.  
In light of this, two possibilities arise in order to generate the baryon asymmetry within this framework.
The first is that a large baryon asymmetry (much larger than the observed amount) is generated in the early Universe and gives rise to the correct asymmetry after entropy dilution. The second possibility is that the decay of the modulus itself generates the asymmetry at temperatures around 10 MeV.  

The second possibility requires baryon number violating decays, since the modulus is a gauge singlet with vanishing baryon number and only couples gravitationally to all SM fields. It is not clear at present if this possibility could naturally occur within a string framework. On the other hand, the first possibility is realized quite naturally. It is well known that the Affleck-Dine (AD) mechanism can generate a large (even ${\cal O}(1)$) baryon asymmetry in a robust manner \cite{Affleck:1984fy, Dine:1995kz}. In particular, this can happen via $B$ and $L$-violating flat-directions in the MSSM denoted by $\Phi$ in general. Within our framework, these flat-directions are also displaced from their late-time minima during inflation, just like the light moduli. The subsequent coherent oscillation and decay of these flat-directions could then generate a baryon asymmetry. In the simple MSSM models realizing this possibility, however, there are two issues. First, as mentioned before, there is the danger of producing \emph{too much} baryon asymmetry and second, the origins of the baryon asymmetry and the DM abundance seem to be decoupled from each other. The presence of light moduli in our framework can provide a resolution to both these issues, in the following manner.

The essential point is that the decay of the lightest modulus generates the LSP abundance (see section \ref{wimp-dm}), and at the same time provides the dilution factor for computing the final asymmetric baryon abundance, thereby relating the two. A careful analysis then gives rise to the following ratio for the two abundances \cite{Kane:2011ih}: \ba \label{BLSPratio} \frac{\Omega_B}{\Omega_{\chi}} \simeq {\cal O}(1)\,\frac{m_{proton}\,m_{pl}\,T_R^2\,\langle\sigma v\rangle}{m_{3/2}\,m_{\chi}}\,\left(\frac{\Phi_0}{X_0}\right)^2. \ea Here $\Phi_0$ and $X_0$ denote the initial displacements of the flat-direction and the lightest modulus during inflation, respectively. Their ratio  above arises in the ratio of the corresponding energy densities and determines how much baryon asymmetry is left after the dilution. Furthermore, as shown in \cite{Kane:2011ih}, flat directions corresponding to the highest dimension operators in the MSSM (which yield the largest $\Phi_0$) naturally give rise to $\frac{\Phi_0}{X_0}$ in the range $10^{-3}-10^{-2}$. Then, for natural values of other quantities in (\ref{BLSPratio}) arising within the framework and consistent with other constraints, such as  $m_{3/2}$ in the 20-100 TeV range, giving rise to $T_R$ around few to 100 MeV (this depends on the modulus decay constant), $m_{\chi}$ around 200 GeV, and $\langle\sigma v\rangle$ around ${\rm few} \times 10^{-6}\,{\rm GeV}^{-2}$, the above ratio is close to the observed value. Note that $\Omega_{\chi}$ in (\ref{BLSPratio}) is \emph{not} the full DM abundance since axions also contribute to the DM abundance. Therefore, the ratio $ \frac{\Omega_B}{\Omega_{\chi}}$ has to be somewhat larger than $0.2$.

\section{Comments and Outlook}\label{conclude}

In this review, we have outlined the typical or generic predictions of a string/$M$ theory vacuum \emph{given} our Universe is a solution of string/$M$ theory with low energy
supersymmetry and grand unification.
We have carefully laid out the broad set of working assumptions under which the results are valid, in section \ref{assumptions}. In addition to the requirement of stabilizing all moduli in a vacuum which solves the gauge hierarchy problem with supersymmetry, these essentially amount to assuming that the supergravity approximation is valid, the Hubble parameter during inflation (or whatever solves the horizon and flatness problems in the early Universe) is larger than $m_{3/2}$, and that the visible sector is weakly coupled until a high scale like the unification scale. Then, many broad predictions can be made for beyond-the-SM physics. For a detailed summary of the results obtained, see section \ref{summary}. In this section, we comment on a few issues.

It is worth addressing complaints which critical readers might have about the whole approach. For example, some may complain that the approach considered here has not tackled any of the deep fundamental problems, such as the understanding of the cosmological singularity at very early times, or the solution of the horizon and flatness problems in the early Universe, or the extremely tiny value of the cosmological constant. On a more mundane but technical level, others may complain that although moduli stabilization has been understood at the effective supergravity level, explicit compactifications with the required properties to stabilize all moduli, \emph{and} a realistic matter and gauge spectrum (such as that of the MSSM), \emph{and} a realistic texture of Yukawa couplings for the quarks, leptons and neutrinos, do not exist. 

Section \ref{philo} addresses the above questions. It is clear that our understanding of these deeper issues is rudimentary at best. However, our main assumption is that our Universe is a solution of string/$M$ theory. If this assumption is correct, then there must be mechanisms present in the theory (albeit unknown to us) which would have solved the first two fundamental  issues at very early times and presumably at very high scales. Our focus in this work is on the broad features of beyond-the-SM physics which depend on our understanding of the Universe at much later times, essentially from around the time of BBN to the present time. Hence, these features are largely decoupled from the first two issues. The cosmological constant, on the other hand is a fundamental problem which persists even at late times. So, regarding the cosmological constant our philosophy is as follows. We only require that the cosmological constant approximately vanishes, with the implicit assumption that the (unknown) mechanism which gives rise to the extremely tiny value of the cosmological constant has no bearing on BSM particle physics. This appears to be a rather conservative assumption since there is no known particle physics process whose outcome depends on the \emph{precise} value of the cosmological constant. 

For the technical complaints it is worth noting that explicit compactifications which stabilize all moduli by incorporating the underlying physical ideas exist for Type IIB compactifications \cite{arXiv:1003.1982}.  For $M$ theory compactifications, although explicit manifolds with such properties do not exist yet, dualities from other corners of string theory suggest that essentially the same mechanism should go through for these compactifications. Similarly, many explicit compactifications realizing a realistic matter and gauge spectrum such as that of the MSSM have been constructed in various corners of string theory \cite{Braun:2005nv}.  While explicit string compactifications realizing \emph{all} these features in a \emph{single} vacuum may have not yet been constructed, the fact that these features exist (separately) in a large class of vacua lends support to the expectation that there should exist 4D string/$M$ theory vacua in the landscape realizing all these features. 

The approach we have espoused is very useful even if predictions do \emph{not} agree with data, since depending upon the nature of experimental data, it could provide insights as to which of the assumptions need to be relaxed. Let us explain this with a few examples. 

One way in which some of the above conclusions could be modified is if the matter and gauge spectrum below the compactification scale is more extended than that of the MSSM. This could give rise to a different prediction for the Higgs mass as well as its properties in general. For example, this could happen if the Higgs couples to additional particles through Yukawa or gauge interactions.  Smilarly,  if FERMI or AMS-02 do not see a WIMP signal in the near future, then this would mean that the component of Dark Matter in the form of WIMPs annihilating to SM states is much smaller. This could  happen for a number of reasons, such as if the stabilizing symmetry for the LSP (like $R$-parity) is violated sufficiently strongly,  or if the LSP resides in a hidden sector very weakly coupled to us so that the lightest superpartner in the visible sector decays to the LSP in the hidden sector, or if the modulus decay constant is smaller than what is theoretically expected (see section \ref{wimp-dm}). The precise form of the data could then distinguish between the different possibilities.

As a final example, if squarks and sleptons are observed at the LHC, this would be in contradiction with some of the basic assumptions of the framework. This would imply one of four possibilities -- a) the moduli potential is very non-generic which makes  \emph{all} moduli masses much larger than $m_{3/2}$, b) the moduli masses are close to $m_{3/2}$ but the Hubble parameter during inflation is $\lesssim m_{3/2}$, so that the moduli are not displaced from their late-time minima, c) the moduli masses are close to $m_{3/2}$ and the Hubble parameter during inflation is larger than $m_{3/2}$, but there  exists a period of thermal inflation at late-times to dilute the entropy production from the decay of moduli \cite{Lyth:1995ka}, or d) the moduli masses are close to $m_{3/2} \gtrsim 30$ TeV, but the squark and slepton masses are also suppressed relative to $m_{3/2}$ in a phenomenologically consistent way. 
So, if experiments observe squark and slepton masses, we learn that our vacuum is non-generic in the  string/$M$ theory framework. 

If the predictions of this framework agree with data on the other hand, it would be an extremely important step in connecting string/$M$ theory to the real world and would open up more opportunities for learning about the string vacuum we live in. 

\acknowledgements{We would like to thank E. Kuflik, R. Lu, A. Pierce, J. Shao, L. Wang, S. Watson, B. Zheng and particularly K. Bobkov for many helpful discussions. P.K. would like to thank the Michigan Center for Theoretical Physics and Yale University for hospitality during part of the work. B.S.A gratefully acknowledges the support of the
STFC, UK. The work of G.K. is supported by the DoE Grant DE-FG-02-95ER40899 and 
and by the MCTP. The work of PK is supported by the DoE Grant DE-FG02-92ER40699.}

\end{document}